\documentclass[11pt]{article}
\usepackage{fullpage}
\usepackage{amssymb, amsthm, amsmath}
\usepackage{doublespace}
\usepackage{graphicx}
\usepackage[authoryear]{natbib}
\usepackage{bm}
\usepackage{verbatim}
\usepackage{times}

\newcommand{\btheta}{ \mbox{\boldmath $\theta$}}
\newcommand{\bmu}{ \mbox{\boldmath $\mu$}}

\newcommand{\bbeta}{ \mbox{\boldmath $\beta$}}
\newcommand{\bdelta}{ \mbox{\boldmath $\delta$}}

\newcommand{\bvarepsilon}{ \mbox{\boldmath $\varepsilon$}}
\newcommand{\bomega}{ \mbox{\boldmath $\omega$}}

\newcommand{\bGamma}{ \mbox{\boldmath $\Gamma$}}

\newcommand{\bA}{ \mbox{\bf A}}

\newcommand{\bx}{ \mbox{\bf x}}
\newcommand{\bX}{ \mbox{\bf X}}
\newcommand{\bB}{ \mbox{\bf B}}
\newcommand{\bZ}{ \mbox{\bf Z}}

\newcommand{\bY}{ \mbox{\bf Y}}
\newcommand{\bz}{ \mbox{\bf z}}

\newcommand{\bT}{ \mbox{\bf T}}
\newcommand{\bI}{ \mbox{\bf I}}
\newcommand{\bs}{ \mbox{\bf s}}
\newcommand{\bb}{ \mbox{\bf b}}

\newcommand{\bV}{ \mbox{\bf V}}

\newcommand{\bC}{ \mbox{\bf C}}

\newcommand{\bR}{ \mbox{\bf R}}

\newcommand{\bW}{ \mbox{\bf W}}

\newcommand{\bF}{ \mbox{\bf F}}
\newcommand{\iid}{\stackrel{iid}{\sim}}
\newcommand{\indep}{\stackrel{indep}{\sim}}
\newcommand{\calR}{{\cal R}}

\newcommand{\calD}{{\cal D}}
\newcommand{\calS}{{\cal S}}

\newcommand{\calX}{{\cal X}}
\newcommand{\calY}{{\cal Y}}
\newcommand{\calZ}{{\cal Z}}

\newcommand{\Matern}{ \mbox{Mat$\acute{\mbox{e}}$rn}}

\newcommand{\beq}{ \begin{equation}}
\newcommand{\eeq}{ \end{equation}}
\newcommand{\beqn}{ \begin{eqnarray}}
\newcommand{\eeqn}{ \end{eqnarray}}

\begin{document}

\begin{center}
{\Large A spectral adjustment for spatial confounding}\\\vspace{6pt}
{\large Yawen Guan\footnote{University of Nebraska - Lincoln}, Garritt L. Page\footnote{Brigham Young University}, Brian J Reich\footnote{North Carolina State University}, Massimo Ventrucci\footnote{University of Bologna} and Shu Yang$^3$} 
\end{center}

\begin{abstract}\begin{singlespace}
\noindent Adjusting for an unmeasured confounder is generally an intractable problem, but in the spatial setting it may be possible under certain conditions.  In this paper, we derive necessary conditions on the coherence between the treatment variable of interest and the unmeasured confounder that ensure the causal effect of the treatment is estimable.  We specify our model and assumptions in the spectral domain to allow for different degrees of confounding at different spatial resolutions.  The key assumption that ensures identifiability is that confounding present at global scales dissipates at local scales.  We show that this assumption in the spectral domain is equivalent to adjusting for global-scale confounding in the spatial domain by adding a spatially smoothed version of the treatment variable to the mean of the response variable.  Within this general framework, we propose a sequence of confounder adjustment methods that range from parametric adjustments based on the $\Matern$ coherence function to more robust semi-parametric methods that use smoothing splines.  These ideas are applied to areal and geostatistical data for both simulated and real datasets. \vspace{12pt} 
\end{singlespace} \end{abstract} 
\newpage

\section{Introduction}\label{s:intro}

A fundamental task in environmental and epidemiological applications is to use spatially-correlated observational  data to estimate the effect of a treatment variable on a response variable.  A key assumption needed to endow an analysis with a causal interpretation is that all relevant confounding variables have been included in the statistical model.  This no-missing-confounders assumption is generally impossible to verify, but in the spatial setting it may be possible to remove the effects of unmeasured confounding variables that are smooth spatial functions by including spatially-correlated random effects in the model.  It is known that regressions with and without spatial random effects can give dramatically different results 
\citep{reich2006effects, paciorek2010importance, hodges2010adding, page:2017, khan&calder:2020}, a phenomena known as spatial confounding. However, it remains unclear how to specify assumptions and methods that lead to a valid causal interpretation in the presence of an unmeasured spatial confounding variable.  In this paper, we propose new methods to adjust for missing spatial confounding variables using spectral methods. 

One explanation of the differences between regression models that do and do not include a spatial random effect is that the treatment variable is collinear with an unmeasured spatial confounding variable.   To account for spatial confounding, \cite{reich2006effects}, \cite{hughes2013dimension} and \cite{prates2019} restrict the residual process to be orthogonal to the treatment process (an approach referred to as restricted spatial regression, RSR). However, RSR makes strong assumptions and can perform poorly when the model is misspecified \citep{hanks2015restricted}. Indeed, in recent work, \cite{khan&calder:2020} showed that the variants of RSR often times are inferior to regression models that completely ignore spatial dependence when coefficient estimation is of interest.  Therefore, proper adjustment for spatial confounding continues to be an important and open problem as evidenced by the many contexts in which it is being addressed (\citealt{hefley_2017, pereira_2020, azevedo2020alleviating, widemberg:2020}).

Alternatives to RSR have appeared in the literature.  For example, \cite{dupont2020spatial} use a two-stage thin plate spline model for covariate and unmeasured spatial confounder. Alternatively,  \cite{thaden&kneib:2018} and \cite{schnell2019mitigating} propose jointly modeling the spatial structure in response and covariate.  The former carries this out through a structural equation model while the later uses a Gaussian Markov random field construction. Both these methods have connections with causal inference (see \citealt{reich2020review} for a recent review of spatial causal inference). In the spatial causal effect setting, \cite{osama19a} permit the spatial causal effect to vary across space.  
In general, adjusting for a missing confounding variable is impossible without further information or assumptions, leading to a variety of approaches and increased attention dedicated to this problem.

We propose to couch spatial regression with missing confounding variables in the spectral domain.
Following \cite{paciorek2010importance} and \cite{Keller_2020}, we focus on the spatial scales of the treatment and missing confounder variable.  We posit a joint model for these variables in the spectral domain \citep[as in][for a prediction problem]{reich2014spectral} and study their coherence, i.e., their correlation at different spatial scales.  (As an aside, \cite{stokes2017study} and \cite{faes:2019} consider a frequency-domain measure of causality from a temporal perspective, although their resulting estimators are quite different than those proposed here.)  The resulting causal effect estimate reveals that the optimal confounder adjustment is a function of the coherence function.  We then show that the optimal confounder adjustment is not estimable without further assumptions, and so we assume that the coherence dissipates at high frequencies, i.e., that there are no unmeasured confounder variables when considering only local spatial variation.  We show this assumption in the spectral domain implies that the model in the spatial domain should include a spatially-smoothed version of the treatment variable as a covariate to adjust for global-scale spatial confounding.  We then develop parametric and nonparametic methods to approximate the optimal confounding adjustment and causal effect while accounting for uncertainty in this approximation. We consider both areal and point-referenced data for Gaussian and non-Gaussian responses. 


\section{Continuous-space modeling framework}\label{s:cont}
In this section, we develop a framework for spatial causal inference for data observed over a continuous spatial domain, and study the identifiability of the causal effect under different confounding scenarios.  Let $X(\bs)$ and $Z(\bs)$ be the observed treatment and confounder processes respectively at the spatial location $\bs \in D\subset\calR^2$ and let vectors $\bX = (X_1,\dots, X_n)^T$ and $ \bZ =  (Z_1,\dots, Z_n)^T$ be the process evaluated at the set of locations $ \calS = \{\bs_1,\dots, \bs_n\} \in D$, where $X_i = X(\bs_i)$ and $Z_i = Z(\bs_i)$. For simplicity we consider only a single treatment and confounding variable, but the results extend to the multivariate setting (see Supplemental Section \ref{s:app:multivariate}). 

Following the commonly used spatial regression model, we assume a linear addictive relationship for the response $\bY = (Y_1,\dots, Y_n)^T$,  \begin{equation}\label{e:spatial_regression}
  \bY = \beta_0 + \beta_x\bX + \beta_z\bZ + \bvarepsilon,
\end{equation}
where $\bvarepsilon=(\varepsilon_1,...,\varepsilon_n)^T$ and $\varepsilon_i\iid\text{Normal}(0,\sigma^2)$. The regression coefficient $\beta_x$ has a causal interpretation under the potential outcomes framework and the stable-unit-treatment-value, consistency, and conditional-treatment-ignorability assumptions (Supplemental Section \ref{s:app:PO}). If we observe the confounder $\bZ$, then identification and estimation of $\beta_x$ is straightforward using multiple linear regression. However, we assume that $\bZ$ is an unmeasured confounder, making $\beta_x$ not identifiable in general.  We propose to exploit the spatial structure of $\bZ$ to mitigate the effects of the unobserved confounder. In Section \ref{s:spec} we propose assumptions in the spectral domain that identify the causal effect in the presence of an unmeasured spatial confounder. 

\subsection{Spectral representation of confounding and identification}\label{s:spec}

We model the dependence between $X(\bs)$ and $Z(\bs)$ using their spectral representations. This allows for different dependence at different spatial scales as each frequency corresponds to a spatial scale, with low frequency corresponding to large spatial scale phenomenon and vise versa.  We assume both processes $X(\bs)$ and $Z(\bs)$ have mean zero and are stationary, and thus have spectral representations
\begin{eqnarray*}
X(\bs) &=& \int \exp(i\bomega^T\bs)\calX(\bomega)d\bomega \\
Z(\bs) &=& \int \exp(i\bomega^T\bs)\calZ(\bomega)d\bomega ,
\end{eqnarray*}
where $\bomega \in \calR^2$ is a frequency. The spectral processes $\calX(\bomega)$ and $\calZ(\bomega)$ are Gaussian with $\mbox{E}\{\calX(\bomega)\}=\mbox{E}\{\calZ(\bomega)\}=0$ and are independent across frequencies, so that for any $\bomega\ne\bomega'$, $\mbox{Cov}\{\calX(\bomega),\calX(\bomega')\} = \mbox{Cov}\{\calZ(\bomega),\calZ(\bomega')\} = \mbox{Cov}\{\calX(\bomega), \calZ(\bomega')\} = 0$. At the same frequency, the covariance of the joint spectral process has the form
\begin{equation}\label{e:Covxz}
    \mbox{Cov}\begin{pmatrix} 
\calX(\bomega) \\
\calZ(\bomega) 
\end{pmatrix} = \begin{pmatrix} 
\sigma_x^2f_{x}(\bomega) & \rho\sigma_x\sigma_zf_{xz}(\bomega) \\
\rho\sigma_x\sigma_zf_{xz}(\bomega) & \sigma_z^2f_{z}(\bomega) 
\end{pmatrix},
\end{equation}
where $\sigma_x^2$ and $\sigma_z^2$ are variance parameters, $f_x(\bomega)>0$ and $f_z(\bomega)>0$ are spectral densities that determine the marginal spatial correlation of $X(\bs)$ and $Z(\bs)$, respectively, and the cross-spectral density $f_{xz}(\bomega)$ determines the dependence between the spectral processes at different frequencies. 

Normalizing the cross-spectral density by each marginal standard deviation, we can derive the coherence function which determines the correlations between the two spectral processes at different frequencies,
\begin{equation}\label{e:coherence}
    \gamma(\bomega) = \rho\frac{f_{xz}(\bomega)}{\sqrt{f_{x}(\bomega)f_{z}(\bomega)}}\in[-1,1].
\end{equation} 
The parameter $\rho$ is a scalar that controls the overall strength of cross-correlation. The coherence function is useful to describe the relationships between  processes \citep{kleiber2017coherence}.

Returning to the response model \eqref{e:spatial_regression}, we let $Y(\bs) = \int \exp(i\bomega^T\bs)\calY(\bomega)d\bomega$ be the spectral representation of the response.  The conditional distribution of $\calY(\bomega)$ given $\calX(\bomega)$, marginalizing over $\calZ(\bomega)$ is 
\begin{eqnarray}\label{e:ygivenx}
   \calY(\bomega)|\calX(\bomega) &\indep& \mbox{Normal}\left(\beta_x \calX(\bomega) + \beta_z\alpha(\bomega)\calX(\bomega),\tau^2(\bomega)+\sigma^2\right) \\
   \alpha(\bomega) &=& \rho\frac{\sigma_zf_{xz}(\bomega)}{\sigma_xf_{x}(\bomega)}=\frac{\sigma_z\sqrt{f_{z}(\bomega)}}{\sigma_x\sqrt{f_{x}(\bomega)}}\gamma(\bomega)\nonumber\\
   \tau^2(\bomega) &=& \beta_z^2\sigma_z^2f_z(\bomega)\left[1-\rho^2\frac{f_{xz}(\bomega)^2}{f_x(\bomega)f_z(\bomega)}\right].
   \nonumber
\end{eqnarray}
The regression coefficient for $\calX(\bomega)$ is $\beta(\bomega) = \beta_x+\beta_z\alpha(\bomega) \ne \beta_x$. The additional term ${\hat \calZ}(\bomega) = \mbox{E}[\calZ(\bomega)|\calX(\bomega)] = \alpha(\bomega)\calX(\bomega)$ is a result of attributing the effect of the unmeasured confounder on the response to the treatment variable, potentially inducing bias in estimating $\beta_x$. 

Therefore, the causal effect $\beta_x$ is identified only if the projection operator $\alpha(\bomega)$ can be assumed to be known or estimated for some prespecified $\bomega$. Of course, $\alpha(\bomega)$ is generally not known and cannot be estimated without further assumptions because $Z(\bs)$, and therefore $\calZ(\bomega)$, is not observed.  We consider two approaches for identification: unconfoundedness at high frequencies so that high-frequency terms identify the causal effect (Section \ref{subsub1}) or a parsimonious and parametric model with constraints on the parameters to ensure identification (Section \ref{subsub2}).

\subsubsection{Unconfoundedness at high-frequencies}\label{subsub1} If we assume that $\alpha(\bomega)\rightarrow 0$ for large $||\bomega||$ then $\mbox{E}\{\calY(\bomega)|\calX(\bomega)\} \approx \beta_x\calX(\bomega)$ and thus $\beta_x$ is identified.  The assumption that $\alpha(\bomega)\rightarrow 0$ for large $||\bomega||$ implies that the cross-spectral density decreases to zero faster than the spectral density of $X$, implying a decrease in confounding in higher frequency or local variations. 
High frequency terms provide the most reliable information about the causal effect because they correlate local changes in the treatment with local changes in the response. An extreme case of local information about the causal effect is the difference in the response for two nearby sites with different treatment levels.  
This local difference eliminates problems caused by omitted variables that vary smoothly over space.  Of course, this cannot completely rule out missing confounding variables that co-vary with both the treatment and the response at high frequencies, but it does lessen the likelihood of spurious confounding effects. 


\subsubsection{Parsimonious coherence model}\label{subsub2} The coherence in (\ref{e:coherence}) simplifies to the constant function $\gamma(\bomega)=\rho C \in (-1,1) $ if we assume $f_{xz}(\bomega) = C \sqrt{f_{x}(\bomega)f_{z}(\bomega)}$ for a constant $C$.  Generalizing the use of a term from \cite{gneiting2010matern}, we refer to this as the parsimonious coherence model.  This imposes the assumption that the correlation between the treatment and missing confounder is frequency-invariant, and this greatly simplifies estimation because the model involves only two spectral densities that can be estimated using the marginal spatial covariances of the response and treatment variables, as described below.  The expression in (\ref{e:ygivenx}) simplifies under the parsimonious model to 
\begin{equation}\label{e:ygivenx:pars}
\calY(\bomega)|\calX(\bomega) \indep \mbox{Normal}\left\{\left(\beta_x +\rho C\beta_z\frac{\sigma_z\sqrt{f_{z}(\bomega)}}{\sigma_x\sqrt{f_{x}(\bomega)}}\right)\calX(\bomega),\left(1-(\rho C)^2
\right)\beta_z^2\sigma_z^2f_z(\bomega)+\sigma^2\right\}.
 \end{equation}
Clearly $\beta_z$ and $\sigma_z$ cannot be individually identified as only their product appears in (\ref{e:ygivenx:pars}), so we set $\beta_z=1$.   Given this restriction and the assumption that $f_x(\bomega)\ne f_z(\bomega)$ for some $\bomega$, Supplemental Section \ref{s:app:IDparism} shows that the remaining parameters $\beta_x$, $\rho C$, $\sigma_x$ and $\sigma_z$ and functions $f_x$ and $f_z$ are all identified.

\subsection{Spatial representation of confounding and identification}\label{s:spatialrep}
Returning to the spatial domain, the response process can be written as 
\begin{eqnarray*}
Y(\bs)|X(\bs), \bs \in \calD &=& \beta_0 + \beta_x X(\bs) + \beta_z{\hat Z}(\bs) + \delta(\bs)\\
{\hat Z}(\bs) &=& \int \exp(i\bomega^T\bs){\hat \calZ}(\bomega)d\bomega = \int \exp(i\bomega^T\bs)\alpha(\bomega)\calX(\bomega)d\bomega,\nonumber
\end{eqnarray*}
and $\delta(\bs)$ is a mean-zero Gaussian process with spectral density $\tau^2(\bomega) + \sigma^2$ independent of $X(\bs)$ and $\hat{Z}(\bs)$.  
If $\alpha(\bomega)$ were known, then ${\hat Z}(\bs)$ would be an appropriate adjustment to the mean to account for the unmeasured confounder. 

The function $\alpha(\bomega)$ acts as a smoothing operator. The oracle (i.e., if $\alpha(\bomega)$ is known) confounder adjustment has the form ${\hat Z}(\bs) = \int K(\bs-\bs') X(\bs') \mathrm{d}\bs'$,
where the kernel function $K(\bs-\bs')$ is the inverse Fourier transform of $\alpha(\bomega)$ (Supplemental Section \ref{s:app:oracle}). Thus the oracle confounder adjustment is conveniently expressed as a kernel-smoothed function of the covariate of interest. It is also straightforward to show that for any $n$ locations $\calS$, 
\begin{equation}\label{e:YgivenX}
    \bY\mid\bX = \beta_0\bm{1} + \beta_x \bX + \beta_z{\hat \bZ}+\bdelta, \mbox{\ \ \ \ where \ \ \ \ }{\hat \bZ} = \Sigma_{zx}\Sigma_x^{-1}\bX,
\end{equation}
$\mbox{Cov}(\bdelta) = \beta_z^2(\Sigma_z-\Sigma_{zx}\Sigma_{x}^{-1}\Sigma_{zx}^T)+\sigma^2I_n$ and $\Sigma_{zx}=\text{Cov}(\bZ,\bX)$, $\Sigma_x = \text{Cov}(\bX)$ and $\Sigma_z = \text{Cov}(\bZ)$. 
The product $\Sigma_{zx}\Sigma_x^{-1}$ serves as a smoothing operator on $\bX$. This representation is convenient for estimation and to determine the strength of confounding dependence between $\bX$ and $\bZ$.  

Including ${\hat \bZ}$ as a covariate in (\ref{e:YgivenX}) effectively removes effects of the large-scale spatial trends in $\bX$ so that the estimate of $\beta_x$ is largely determined by high-frequency terms. This expression also lays bare the importance of assuming that $\alpha(\bomega)$ converges to zero for large frequencies or that $X$ and $Z$ have different spectral densities.  If restrictions are not placed on $\alpha(\bomega)$, then it may be that $\Sigma_{zx}=\Sigma_x$ and thus ${\hat \bZ}=\bX$, giving the non-identifiable model $\bY = \beta_0\bm{1} + \beta_x\bX + \beta_z\bX+\bdelta$.


\section{Continuous-space estimation strategies}\label{s:cont:est}
Below we define two approaches for modeling $\alpha(\bomega)$: a parametric model induced by a bivariate $\Matern$ covariance function for $X(\bs)$ and $Z(\bs)$ (Section \ref{s:cont:matern}), and a semi-parametric model using a spectral mixture prior for $\alpha(\bomega)$ (Section \ref{s:cont:semi}).

\subsection{The bivariate $\Matern$ model}\label{s:cont:matern}

The bivariate $\Matern$ model \citep{gneiting2010matern,apanasovich2012valid} is a flexible parametric model for the spectral densities $f_x$, $f_z$ and $f_{xz}$.  The $\Matern$ spectral density function for a process in two dimensions is $  m(\bomega;\nu,\phi) = \nu \phi^{-2\nu}    
   \left(\phi^{-2}+||\bomega||^2 \right)^{-(\nu+1)}.$
The spectral density is defined by the smoothess parameter $\nu>0$ and the spatial range $\phi>0$.  The bivariate $\Matern$ may have  different parameters for each process, $f_j(\bomega) = m(\bomega;\nu_j,\phi_j)$, for $j\in\{x,z,xz\}$, although constraints on the range and smoothness parameters are needed to ensure that the coherence is positive definite for all $\bomega$ \citep{gneiting2010matern,apanasovich2012valid}.

With this modeling assumption, the projection operator has the form
\begin{equation*}\label{e:alpha_matern}
 \alpha(\bomega) = \rho\frac{\sigma_zm(\bomega;\nu_{xz},\phi_{xz})}{\sigma_xm(\bomega;\nu_x,\phi_x)}.
\end{equation*}
Therefore, if the cross-spectral density decays faster than the covariate spectral density then the confounding adjustment is small for high frequencies.  This expression simplifies further if we assume a common range parameter $\phi_j=\phi$ for $j\in\{x,z,xz\}$, 
\begin{equation}\label{e:alpha_matern:common_range}
 \alpha(\bomega) = \rho\frac{\sigma_z}{\sigma_x}\left(\phi^{-2}+||\bomega||^2\right)^{-(\nu_{xz}-\nu_x)}.
\end{equation}
In this case, we have unconfoundedness at high frequencies, $\alpha(\bomega)\rightarrow 0$, if and only if $\nu_{xz}>\nu_x$, i.e., the cross-covariance is smoother than the covariate covariance.  On the other hand, if we assume a common smoothness $\nu_{j}=\nu$ for $j\in\{x,z,xz\}$, then $\alpha(\bomega)\rightarrow \left(\frac{\phi_{x}}{\phi_{xz}}\right)^{2\nu}$ and confounding persists at high frequencies, regardless of the range parameters.  Therefore, a common range parameterization allows us to identify the causal effect by reducing high-resolution confounding while a common smoothness parameterization will not. For simplicity, we will assume a common range in the remainder of this section.

\begin{figure}
\caption{{\bf Example confounder adjustment for the bivariate $\Matern$}: The covariate process $X(\bs)$ is generated from a $\Matern$ process with range $\phi_x=1$ and smoothness $\nu_x=1$ on a $50\times 50$ grid with grid spacing one, and the plots below give the confounder adjustment ${\hat Z}(\bs)$ for range $\phi_{xz}=\phi_x$ and smoothness $\nu_{xz}=c\nu_x$ for $c\in\{1,3,5\}$ (${\hat Z}(\bs)=X(\bs)$ for $c=1$).}\label{f:Zhat}
\centering
\includegraphics[page=1,width=0.32\textwidth]{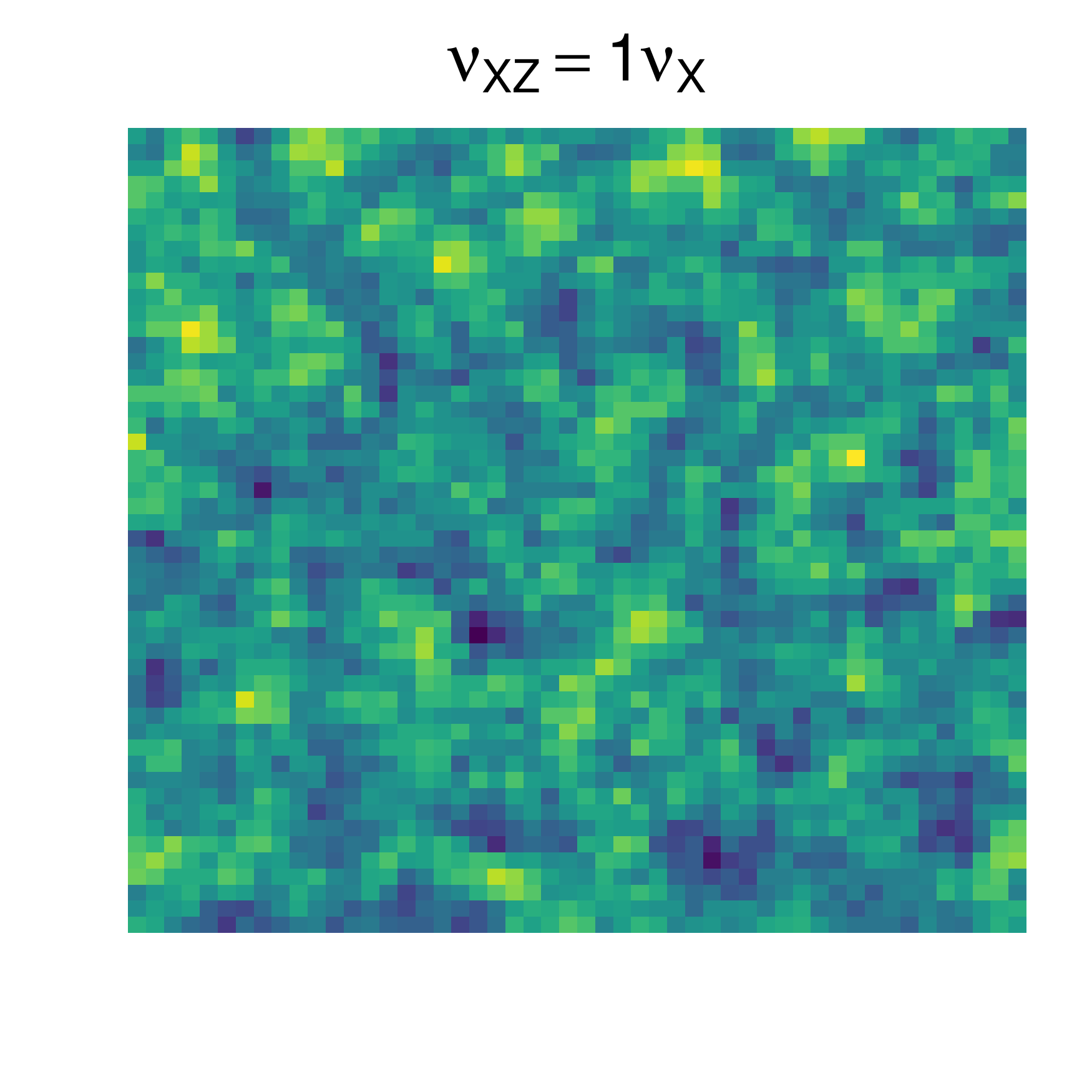}	
\includegraphics[page=2,width=0.32\textwidth]{figs/Matern_Zhat.pdf}	
\includegraphics[page=3,width=0.32\textwidth]{figs/Matern_Zhat.pdf}	
\end{figure}

Figure \ref{f:Zhat} illustrates the confounder adjustment ${\hat \bZ}$ in (\ref{e:YgivenX}) for the bivariate $\Matern$ model
with a common range and different smoothness $\nu_{zx}=c\nu_x$ for increasing values of $c$.  Increasing $c$ implies increasing decay rates of $\alpha(\bomega)$. The original simulated $\bX$ is plotted in the left panel of  Figure \ref{f:Zhat} with $c=1$, in which case we have ${\hat \bZ}=\bX$ and thus a completely confounded model.  In the cases with $c>1$ the confounder adjustment ${\hat \bZ}$ is a smoothed version of $\bX$. Therefore, including ${\hat \bZ}$ as a covariate in the model removes large-scale trends in $\bX$ to adjust for confounding at low-frequencies.  

The unmeasured confounder cannot be observed, making it difficult to estimate all the parameters in the bivariate $\Matern$ model. Therefore, additional constraints are required for estimation of the parameters. Assuming $\beta_z=1$ and a common range parameter,  sufficient conditions (Supplemental Section \ref{s:app:IDmatern}) for identifiability of the remaining parameters $\{\beta_x,\rho,\nu_x,\nu_z,\nu_{xz},\sigma_x,\sigma_z\}$
are a large cross-smoothness parameter $\nu_{xz}>\max\{\nu_x,(\nu_x+\nu_z)/2\}$ and that $\rho^2 < \frac{\nu_x\nu_z}{\nu_{xz}^2}$.

The common range model simplifies further under the parsimonious model in (\ref{e:ygivenx:pars}) with $C = \frac{(\nu_x+\nu_z)/2}{\sqrt{\nu_x\nu_z}}$, $|\rho| < \frac{\sqrt{\nu_x\nu_z}}{(\nu_x+\nu_z)/2}$, and $f_{xz}(\bomega) = m(\bomega;(\nu_x+\nu_z)/2,\phi)$.
This is the parsimonious $\Matern$ model of \cite{gneiting2010matern}, i.e., the cross smoothness equals the average of the marginal smoothness parameters, $\nu_{xz}=(\nu_x+\nu_z)/2$. Under this model the confounder adjustment becomes
\begin{equation}\label{e:alpha_matern:parsi}
 \alpha(\bomega) = \rho\frac{\sigma_z}{\sigma_x}\left(\phi^{-2}+||\bomega||^2\right)^{-(\nu_{z}-\nu_x)/2}
\end{equation}
and thus $\alpha(\bomega)\rightarrow 0$ if and only if $\nu_{z}>\nu_x$, i.e., the missing confounder is smoother than the treatment.  On the other hand, if $\nu_z<\nu_x$ then $\alpha(\bomega)\rightarrow \infty$, which is clearly undesirable. 


\subsection{Semi-parametric model}\label{s:cont:semi}
Rather than indirectly modeling the projection operator $\alpha(\bomega)$ via a model for the cross-covariance function, in this section we directly model $\alpha(\bomega)$ using a flexible mixture model. We use a linear combination of B-splines as follows:
\begin{equation*}
	\alpha( \omega) = \sum_{l=1}^LB_l( \omega )b_l, \hspace{2em} 0<\omega<\pi/\bigtriangleup_s
\end{equation*}     
where $ \omega = ||\bomega|| $, $ B_l(\omega) $ are B-spline basis functions, $ b_l $ are the associate coefficients and $\bigtriangleup_s=\text{max}\left\lbrace ds_1, ds_2 \right\rbrace $ for grid spacing $(ds_1, ds_2)$. A uniform sequence of knots $\{\omega_1^\ast,...,\omega_L^\ast\}$ are placed to cover the interval $[0,\pi/\bigtriangleup_s]$, such that $ 0\in (\omega_1^\ast, \omega_2^\ast) $ and $ \pi/\bigtriangleup_s\in (\omega_{L-1}^\ast, \omega_L^\ast) $. The interval upper bound $\pi/\bigtriangleup_s$ is the largest spectrum that can be observed from uniformly spaced data due to aliasing \citep{fuentes2010spectral}.  We have restricted the projection operator to be isotropic by letting $ \omega = ||\bomega|| $, but this can be relaxed by using bivariate spline functions. Other mixture priors \citep{reich2012nonparametric, jang2017scalable,chen2018spectral} can also be used for modeling the projection operator

The B-spline mixture model for $ \alpha(\omega) $ does not have a closed-form inverse Fourier transformation. We approximate the kernel-smoothed function $ K(\bs -\bs') $ with a finite sum at a set of equally spaced frequencies $\mathcal{F}=\left\lbrace \omega^f_{1} ,\dots, \omega^f_m  \right\rbrace $ with spacing $ \bigtriangleup_{\mathcal{F}} $ and $ \omega^f_m = \pi/\bigtriangleup_s$ following \cite{Qadir_Sun_2020}
\begin{equation*}
	K(\bs-\bs') = \sum_{\omega^f\in\mathcal{F}}   h\left( \frac{2\pi\omega^f}{h}\right) ^{\kappa+1} \mathcal{J}_\kappa(\omega^f h) \alpha(\omega^f)\bigtriangleup_{\mathcal{F}},
\end{equation*} 
where $ h =||\bs-\bs'||$, $ \kappa =d/2 - 1$ and $ \mathcal{J}_\kappa(\cdot) $  is a Bessel function of the first kind of order $ \kappa $ \citep{watson1995}. This approximation allows us to directly compute confounder adjustment in the spatial domain, which would otherwise require Fourier transform of data to perform analysis in the spectral domain. The confounder adjustment is then given by \begin{eqnarray}
	{\hat Z}(\bs) &=& \sum_{l=1}^Lb_l{\hat Z}_l(\bs)\nonumber\\
	\label{e:Zhat_j}
	{\hat Z}_l(\bs) &=&  \sum_{\omega^f\in\mathcal{F}}  (2\pi\omega^f)^{\kappa+1} B_l(\omega^f)\bigtriangleup_{\mathcal{F}} \int  \frac{\mathcal{J}_\kappa(\omega^f h)}{h^\kappa} X(\bs') d\bs'.
\end{eqnarray} 
When $X(\bs)$ is observed on a grid, the integral can be approximated $\int  \frac{\mathcal{J}_\kappa(\omega^f h)}{h^\kappa} X(\bs') d\bs'  = \frac{1}{n}\sum_{i=1}^n \frac{\mathcal{J}_\kappa(\omega^f h_i)}{h_i^\kappa} X(\bs_i)$ with $h_i=||\bs-\bs_i||$. For non-gridded data, the covariate can be interpolated to a grid and this discrete approximation to the grid can be applied.  The confounder adjustment covariates ${\hat Z}_l(\bs)$ are precomputed to reduce computation during model fitting. We then fit spatial model
\begin{equation}\label{e:cont:Y_Zhat}
	Y(\bs) = \beta_0+\beta_xX(\bs)+\sum_{l=1}^L b_l\hat{Z}_l(\bs) + \delta(\bs) + \epsilon(\bs),
\end{equation}
where $\beta_z=1$ for identification and $\delta(\bs)$ is modeled as a Gaussian process. The coefficients $(b_1,...,b_L)^T$ are given intrinsic autoregressive priors with full conditional distributions $b_k|b_{(-k)} \sim \mbox{Normal}({\bar b}_{k}, \sigma_b^2/N_k)$ where ${\bar b}_{k}$ is mean of the $N_k$ coefficients $b_l$ with $|l-k|=1$ (so $N_1=N_L=1$ and $N_2=...,N_{L-1}=2$).

\section{Discrete-space methodology}\label{s:discrete}

The concepts developed in Sections \ref{s:cont} and \ref{s:cont:est} for the continuous spatial domain naturally extend to the discrete case with the spatial domain comprised of $n$ regions. For region $i$, let $Y_i$, $X_i$ and $Z_i$ be the response, treatment and confounding variables, respectively, and $\bY=(Y_1,...,Y_n)^T$, $\bX=(X_1,...,X_n)^T$ and $\bZ=(Z_1,...,Z_n)^T$.  Assuming $\bZ$ is known and the same causal assumptions as in Section \ref{s:cont}, the structural model
\begin{equation}\label{e:obs_outcomes}
  \bY = \beta_0\bm{1} + \beta_x\bX + \beta_z\bZ + \bvarepsilon,
\end{equation}
permits a causal intepretation of $\beta_x$. If we observe the confounder $\bZ$, then estimation of the causal effect $\beta_x$ is straightforward using multiple linear regression.  However, we assume that $\bZ$ is an unmeasured confounder.   

\subsection{A spectral model for confounding}

Since we do not observe $\bZ$, we assume that it follows a spatial model.  In the discrete case, dependence between the regions is often described by a region adjacency structure.  Let $a_{ij}=1$ if regions $i$ and $j$ are adjacent and 0 otherwise, and $m_i$ be the number of regions adjacent to region $i$. A common model based on this adjacency structure is the conditionally autoregressive (CAR) model \citep{gelfand2010handbook} using the Leroux parameterization \citep{leroux2000estimation}.  For an arbitrary vector $\bz=(z_1,...,z_n)^T$, the CAR model is \begin{equation}\label{e:theta}
 \bz\sim\text{Normal}\left\{\bmu,\sigma_z^2[(1-\lambda){\bf I}_n+\lambda_z\bR]^{-1}\right\},
\end{equation}
where $\bmu=(\mu_1,...,\mu_n)^T$ is the mean vector, $\sigma_z^2$ determines the overall variance, $\lambda_z\in[0,1]$ controls the strength of spatial dependence (larger $\lambda_z$ gives stronger dependence) and $\bR$ is an $n\times n$ matrix with $(i,j)$ off-diagonal element $-a_{ij}$ and $i^{th}$ diagonal element $m_i$.  We denote this model as $\bz\sim\text{CAR}(\bmu,\sigma_z^2,\lambda_z)$.


An advantage of the Leroux parameterization is that the spatial covariance can be written as
\begin{equation}\label{e:eigen}
 \sigma_z^2[(1-\lambda_z){\bf I}_n+\lambda_z\bR]^{-1} = \sigma_z^2\bGamma[(1-\lambda_z){\bf I}_n+\lambda_z\bW]^{-1}\bGamma^T
\end{equation}
where the spectral decomposition of $\bR$ is $\bR=\bGamma\bW\bGamma^T$ for orthonormal eigenvector matrix $\bGamma$ and diagonal eigenvalue matrix $\bW$ with $k^{th}$ diagonal element $\omega_k\ge 0$, ordered so that $\omega_1\le...\le \omega_n$.   Assuming all variables have the same adjacency structure $\bR$, this allows us to project the model into the spectral domain using graph Fourier transform \citep{sandryhaila:2013}, $\bY^* = \bGamma^T\bY=(Y^*_1,...,Y^*_n)^T$, $\bX^* = \bGamma^T\bX=(X^*_1,...,X^*_n)^T$ and $\bZ^* = \bGamma^T\bZ=(Z^*_1,...,Z^*_n)^T$. This transformation decorrelates the model and gives 
\begin{equation}\label{e:YstargivenZ}
  Y_k^*|\bX^*,\bZ^* \indep\text{Normal}\left(\beta_0M_k + \beta_xX^*_k + \beta_zZ^*_k, \sigma^2\right),
\end{equation}
where $M_k$ is the sum of the $k^{th}$ column of $\bGamma$ and $(X_k^*,Z_k^*)$ are independent across $k$. To exploit this decorrelation property of the graph Fourier transform, we conduct all analyses of Gaussian data for discrete spatial domain in the spectral scale.

Comparing the discrete and continuous cases, the eigenvalue $\omega_k$ is analogous to frequency, $\bomega$.  Terms with small $\omega_k$ have large variance and measure large-scale trends in the data.  For example, it can be shown that if the $n$ locations form a connected graph, then $\omega_1=0$ and $Y_1^*$ is proportional to the mean of $\bY$.  In contrast, terms with large $\omega_k$ have small variance and represent small-scale features. Using this analogy, in the remainder of this section we extend two of the continuous-domain methods of Section \ref{s:cont:est} to the discrete case.

\subsection{Bivariate CAR model}\label{s:discrete:biCAR}

As in Section \ref{s:cont}, we assume a joint model for $\bX^*$ and $\bZ^*$.  We assume the pairs $(X_k^*,Z_k^*)$ are independent across $k$, and Gaussian with mean zero and covariance
\begin{equation}\label{e:CovxzDiscrete}
    \mbox{Cov}\begin{pmatrix} 
X_k^* \\
Z_k^* 
\end{pmatrix} = \begin{pmatrix} 
\sigma_x^2f_x(\omega_k) & \rho\sigma_x\sigma_zf_{xz}(\omega_k) \\
\rho\sigma_x\sigma_zf_{xz}(\omega_k) & \sigma_z^2f_{z}(\omega_k) 
\end{pmatrix},
\end{equation}
where $\sigma_x^2$ and $\sigma_z^2$ are variance parameters, $f_x(\omega_k)>0$ and $f_z(\omega_k)>0$ are variance functions that determine the covariance of $\bX$ and $\bZ$, respectively, and scalar $\rho$ and function $f_{xz}(\omega_k)$ determine the dependence between $\bX$ and $\bZ$.  For the Leroux CAR model we have $f_j(\omega_k) = 1/(1-\lambda_j+\lambda_j\omega_k)$ for $j\in\{x,z\}$ so that the marginal distributions are $\bX\sim\text{CAR}({\bf 0},\sigma_x^2,\lambda_x)$ and $\bZ\sim\text{CAR}({\bf 0},\sigma_z^2,\lambda_z)$.

One possible parametric cross-covariance model is $f_{xz}(\omega) = 1/(1-\lambda_{xz}+\lambda_{xz}\omega)$. As with the bivariate $\Matern$, the $f_{xz}$ has the same functional form as $f_x$ and $f_z$.  Constraints are required to ensure the covariance in (\ref{e:CovxzDiscrete}) is positive definite, i.e., that
\begin{equation}\label{e:condition} \rho^2(1-\lambda_x+\lambda_xw)(1-\lambda_z+\lambda_zw)<(1-\lambda_{xz}+\lambda_{xz}w)^2
\end{equation}
for all $w\in\{\omega_1,...,\omega_n\}$.
Necessary conditions for (\ref{e:condition}) to hold for all $w\ge 0$ are
$$\rho^2(1-\lambda_x)(1-\lambda_z)<(1-\lambda_{xz})^2 \mbox{\ \ \ \ and \ \ \ \ }
\rho^2\lambda_x\lambda_z<\lambda_{xz}^2,$$
but these conditions are not sufficient and not even necessary when considering only $w\in\{\omega_1,...,\omega_n\}$.

Assuming the covariance parameters give a valid covariance, then marginalizing over $Z_k^*$ and setting $\beta_z=1$ (as in Section \ref{s:cont}) for identification gives
\begin{equation}\label{e:YparsiCAR}
Y_k^*|X_k^*\indep\mbox{Normal}\left(\beta_0M_k + \beta_xX_k^* + \alpha(\omega_k)X_k^*,\tau^2(\omega_k)+\sigma^2\right)
\end{equation}
where $\alpha(\omega_k) = \rho\frac{\sigma_z}{\sigma_x}\frac{1-\lambda_x + \lambda_x \omega_k}{1-\lambda_{xz} + \lambda_{xz} \omega_k}$ and $\tau^2(\omega_k) = \sigma_z^2/(1-\lambda_z+\lambda_z\omega_k)-\rho^2\sigma_z^2\frac{1-\lambda_x + \lambda_x \omega_k}{(1-\lambda_{xz} + \lambda_{xz} \omega_k)^2}$. Therefore, $\alpha(\omega)\rightarrow\rho\frac{\sigma_z\lambda_x}{\sigma_x\lambda_{xz}}$ as $\omega\rightarrow\infty$ and thus the high-resolution confounding effect is smallest when $\lambda_x$ is smaller than $\lambda_{xz}$.

The parsimonious cross-covariance model is $f_{xz}(\omega_k)=\sqrt{f_x(\omega_k)f_z(\omega_k)}$ giving $\mbox{Cor}(X_k^*,Z_k^*)=\rho$ for all $k$.  With this simplification, any $\rho\in(-1,1)$ and $\lambda_x,\lambda_z\in(0,1)$ give a valid covariance and the terms in (\ref{e:YparsiCAR}) reduce to
$\alpha(\omega_k) = \rho\frac{\sigma_z}{\sigma_x}\sqrt{\frac{1-\lambda_x + \lambda_x \omega_k}{1-\lambda_z + \lambda_z \omega_k}}$ and $\tau^2(\omega_k) = \sigma_z^2(1-\rho^2)/(1-\lambda_z+\lambda_z\omega_k)$.  The expression for $\alpha(\omega_k)$ shows that when the missing confounder is smoother than the treatment ($\lambda_z>\lambda_x$) there is less counfounding for high-resolution terms.  However, unlike the parsimonious Mat\'{e}rn model for the continuous-space, the missing confounder need not be smoother than the treatment for identifiability, because for any $\lambda_z$ and $\lambda_x$, the casual effect can be identified as long as $\lambda_x\ne\lambda_z$ as shown in Supplemental Section \ref{s:app:IDparsiCAR}.

In the spatial domain, the parsimonious model is 
\begin{equation}\label{e:CAR:parsi:Y}
 \bY|\bX,\bV\sim\mbox{Normal}
\left(\beta_0{\bf 1}+\beta_x\bX + \bGamma\bA\bGamma^T\bX + \bV,\sigma^2\bI_n\right),
\end{equation}
where $\bV\sim\mbox{CAR}\{{\bf 0},\sigma_z^2(1-\rho^2),\lambda_z\}$  and $\bA$ is diagonal with $k^{th}$ diagonal element $\alpha(\omega_k)$.  The term $\bGamma\bA\bGamma^T\bX$ adjusts for missing spatial confounders and the term $\bV$ captures spatial variation that is independent of $\bX$. In this case with  $\lambda_{z}>\lambda_x$, the confounder adjustment $\bGamma\bA\bGamma^T\bX$ smooths $\bX$ by first projecting into the spectral domain by multiplying by $\bGamma^T$, then dampening the high-frequency terms with large $\omega$ and thus small $\alpha(\omega)$ by multiplying by $\bA$, and finally projecting back in the spatial domain by multiplying by $\bGamma$.

\subsection{Semi-parametric CAR model}\label{s:discrete:semi}

Mirroring Section \ref{s:cont:semi}, rather than specify a parametric joint model for $(X_k^*,Z_k^*)$, we directly specify a flexible model for the confounder adjustment, $\alpha(\omega)$.  The joint model is specified first with the conditional model
$$Z_k^*|X_k^*\indep\mbox{Normal}\left(\alpha(\omega_k) X_k^*,\frac{\sigma_z^2}{1-\lambda_z+\lambda_z \omega_k}\right).$$
In the spatial domain, this implies that $\bZ|\bX\sim\mbox{CAR}(\bGamma\bA\bGamma^T\bX,\sigma_z,\lambda_z)$, where $\bA$ is diagonal with diagonal elements $\{\alpha(\omega_1),...,\alpha(\omega_n)\}$.  Therefore, with any valid marginal distribution of $\bX$, the joint model of $\bX$ and $\bZ$ is well defined.  However, since $\bX$ is observed we do not need a model for its marginal distribution.

Marginalizing over the unknown $\bZ^*$ gives
\begin{equation}\label{e:Ystar2}
  Y_k^*|X_k^* \indep\text{Normal}\left(\beta_0M_k + \beta(\omega_k)X^*_k,\frac{\sigma_z^2}{1-\lambda_z+\lambda_z\omega_k} + \sigma^2\right),
\end{equation}
where $\beta(\omega_k)=\beta_x+\alpha(\omega_k)$. Following Section \ref{s:cont:semi}, we assume that $\alpha(\omega_n)=0$ so that $X_n^*$ and $Z_n^*$ are uncorrelated for the highest-frequency term.  This implies that $\beta(\omega_n)=\beta_x$ and $\mbox{E}(Y_n^*)=\beta_0M_n + \beta_xX_n$, and thus the final term  supplies unbiased information about the true causal effect $\beta_x$. Of course, a single unbiased term is insufficient for estimation, and so we further assume that $\alpha(\omega)$ varies smoothly over $\omega$ to permit semi-parametric estimation of $\beta_x$.     

We fit the model (\ref{e:Ystar2}) with a covariate effect that is allowed to vary with $k$ to separate associations at different spatial resolutions.  Although other smoothing techniques are possible, the frequency-specific coefficients are smoothed using the basis expansion
\begin{equation}\label{e:B}
   \beta(\omega) = \sum_{l=1}^LB_{l}(\omega)b_l,  
\end{equation}
where $B_{l}(\omega)$ are B-spline basis functions and the $b_l$ are the associated coefficients.  Under the assumption that $\alpha(\omega_n)=0$, we use the posterior distribution of $\beta(\omega_n)$ to summarize the causal effect $\beta_x$.

In the spatial domain, the semi-parametric CAR model can be written as (\ref{e:cont:Y_Zhat}), 
\begin{equation}\label{e:CAR:semi:Y2}
 \bY|\bX,\bV\sim\mbox{Normal}
\left(\beta_0{\bf 1}+\sum_{l=1}^L{\hat \bZ}_lb_l + \bV,\sigma^2\bI_n\right),
\end{equation}
with $\bV\sim\mbox{CAR}({\bf 0},\sigma_z^2,\lambda_z)$, ${\hat \bZ}_l = \bGamma\bB_l\bGamma^T\bX$,  $\bB_l$ is the diagonal matrix with spline basis functions, $\{B_l(\omega_1),...,B_l(\omega_n)\}$, on the diagonal and the regression coefficient are modelled as described below (\ref{e:Zhat_j}).  The constructed covariates ${\hat \bZ}_l$ can be precomputed prior to estimation, and thus computation resembles a standard spatial analysis with $L$ known covariates.  As above, under the assumption of no confounding for large $\omega$, we use the posterior of $\beta_x=\sum_{l=1}^LB_{l}(\omega_n)b_l$ to summarize the causal effect.

 

\section{Generalized linear mixed spatial models}\label{s:GLM}

Many of the methods proposed for continuous and discrete domains can be applied to non-Gaussian data using the generalized linear mixed modeling framework. Let $g\{\mbox{E}(Y_i|X_i,Z_i)\}=\theta_i=\beta_x X_i +Z_i$ for link function $g$. Our general approach is to build a joint model for $(X_i,Z_i)$, and then compute the conditional distribution of $\btheta = (\theta_1,...,\theta_n)^T$ given $\bX$.  Using the notation in (\ref{e:YgivenX}), the conditional distribution is 
$$\btheta|\bX \sim 
\mbox{Normal}\left(\beta_0{\bf 1} + \beta_x\bX + {\hat \bZ}, \Sigma_z-\Sigma_{zx}\Sigma_{x}^{-1}\Sigma_{zx}^T\right).$$
For example, the bivariate $\Matern$ could then be fit by estimating the $\Matern$ correlation parameters in ${\hat \bZ}$ and $\Sigma_j$ for $j\in\{x,z,xz\}$.  The semi-parametric model in Section \ref{s:cont:semi} can be fit by assigning $\btheta$ a Gaussian process prior with $\Matern$ covariance function and mean function  $\mbox{E}(\theta_i|X_i)=\beta_0+\beta_xX(\bs_i)+\sum_{l=1}^L\beta_l{\hat Z}_l(\bs_i)$ where ${\hat Z}_l$ is defined in (\ref{e:Zhat_j}).  

The general bivariate CAR model in Section \ref{s:discrete:biCAR} gives
$$\btheta|\bX\sim\mbox{Normal}(\beta_0{\bf 1} + \beta_x\bX + \bGamma\bA\bGamma^T\bX,\bGamma\bT\bGamma^T),$$
where $\bA$ and $\bT$ are diagonal with $k^{th}$ diagonal elements $\alpha(\omega_k)$ and $\tau^2(\omega_k)$, respectively, given below (\ref{e:YparsiCAR}).  This reduces to $$\btheta|\bX\sim\mbox{CAR}(\beta_0{\bf 1} + \beta_x\bX + \bGamma\bA\bGamma^T\bX,(1-\rho^2)\sigma_z^2,\lambda_z)$$ under the parsimonious CAR model.  If we assume the semi-parametric model in Section \ref{s:discrete:semi}, and thus $\beta_x\bX+\bGamma\bA\bGamma^T\bX = \sum_{l=1}^L{\hat \bZ}_lb_l$, then the spatial random effects distribution is 
$$
\btheta|\bX\sim\mbox{CAR}
\left(\beta_0{\bf 1}+ \sum_{l=1}^L{\hat \bZ}_lb_l,\sigma_z^2,\lambda_z\right).
$$

Unlike for Gaussian data, the MCMC cannot proceed in the spectral domain and matrix multiplication $\bGamma\bA\bGamma^T\bX=\bGamma\bA\bX^*$ is required when updating the parameters in $\alpha(\omega_k)$.  For the semi-parametric model much of the computational burden can be shouldered outside the MCMC loop by precomputing matrices as $\bGamma\bA\bGamma^T\bX = \sum_{l=1}^L{\tilde \bX}_lb_l$ where ${\tilde \bX}_l = \bGamma\bB_l\bGamma^T\bX$ and $\bB_l$ is the diagonal matrix with spline basis functions, $\{B_l(\omega_1),...,B_l(\omega_n)\}$, on the diagonal.  After this computation, the method can be fit using standard software for spatial generalized linear models such as {\tt INLA} or {\tt OpenBUGS}, and the posterior of $\beta_x=\sum_{l=1}^LB_l(\omega_n)b_l$ summarizes the causal effect.  As a concrete example for the discrete domain, in Section \ref{s:covid} we consider two applications that require fitting Poisson and Negative-Binomial regression models. 


\section{Simulation study}\label{s:sim}

In this section, we illustrate the proposed methods using simulated data.  The objectives are to illustrate the adverse effects of spatial confounding for standard methods, and to compare parametric and nonparametric alternatives under a range of true coherence functions in terms of bias and coverage of 95\% intervals for the causal effect.  We begin with the simpler discrete case in Section \ref{s:sim_disrete} and then proceed to the continuous case in Section \ref{s:sim_cont}.

\subsection{Discrete-space}\label{s:sim_disrete}

Data are generated at $n$ spatial locations $\bs_1,...,\bs_n$ arranged as the $40\times 40$ square grid with grid spacing one.  The CAR model uses rook neighborhood structure so that $A_{ij} = 1$ if and only if $||\bs_i-\bs_j||=1$.  Rather than simulating data directly from our spectral model, we generate data as \begin{eqnarray}\label{e:sim:data:disc}
\bX &\sim& \text{CAR}\left({\bf 0},\sigma_x^2,\lambda\right),\\  
\bZ|\bX &\sim &\text{CAR}\left(\beta_{xz}\bW\bX,\sigma_z^2,\lambda\right)\nonumber\\  \bY|\bX,\bZ&\sim&\text{Normal}\left(\beta_x\bX+\beta_z\bZ,\sigma^2{\bf I}_n\right)\nonumber
\end{eqnarray}
where $\bW$ is the kernel smoothing matrix with bandwidth $\phi$, i.e., the matrix with $(i,j)$ element $W_{ij} = w_{ij}/(\sum_{l=1}^nw_{il})$ and $\log(w_{ij}) = -(||\bs_i-\bs_j||/\phi)^2$.
Including the kernel-smoothed $\bX$ in the mean of $\bZ$ induces low-resolution dependence between $\bX$ and $\bZ$ while ensuring that $\bZ$ is a smoother process than $\bX$.  In all cases we take $\sigma_x^2=1.7$, $\sigma_z^2=1$, $\lambda=0.95$, $\beta_x=\beta_z=0.5$ and $\sigma^2=0.25^2$, and we
vary the importance of the strength of dependence between $\bX$ and $\bZ$ via $\beta_{xz}\in\{0,1,2\}$, and the kernel bandwidth $\phi\in\{1,2\}$. For each combination of these parameters we generate 500 datasets. A representative dataset for each simulation scenario is plotted in Supplementary Section \ref{s:app:datasets}.

Figure \ref{f:corr_plots} plots the induced correlations in the spectral domain for each scenario with $\beta_{xz}>0$, which is proportional to $\alpha(\omega_k)$.  The correlation is non-zero for only low-frequency terms with small $\omega_k$ when the bandwidth is $\phi=2$, but correlation spills over to high-frequency terms with large $\omega_k$ when $\phi=1$, especially when $\beta_{xz}=2$.  Therefore, the assumption of no confounding at high frequencies is questionable when $\phi=1$, and these scenarios are used to examine sensitivity to this key assumption.  Also, these data-generating scenarios violate the parsimonious assumption that the correlation is constant across frequency, and so they illustrate the effects of mispecifying the parametric model.

\begin{figure}
\caption{{\bf Correlations in the spectral domain for the simulation study}: $\text{Cor}(X^*_k,Z^*_k)$ by the associated eigenvalue $\omega_k$ for different kernel bandwidth ($\phi$) and strength of exposure/confounder dependence ($\beta_{xz}$); in these four designs defined by the real-space average (over locations) $\text{Cor}(X_i,Z_i)$ are 0.62 when $\phi=1$ and $\beta_{xz}=1$, 0.80 when $\phi=1$ and $\beta_{xz}=2$, 0.46 when $\phi=3$ and $\beta_{xz}=1$, and 0.63 when $\phi=3$ and $\beta_{xz}=2$.}\label{f:corr_plots}
\centering
\includegraphics[page=1,width=.4\textwidth]{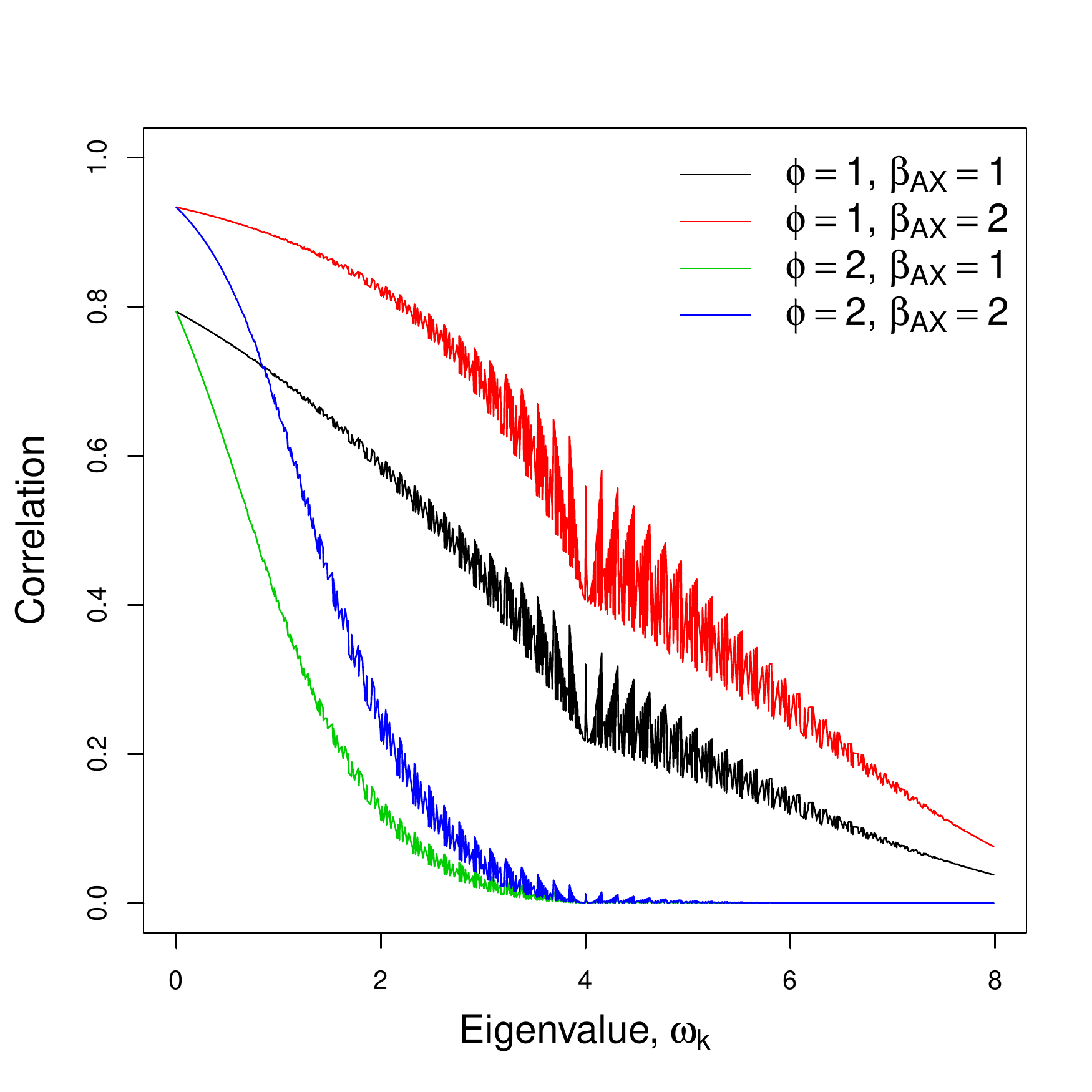}	\end{figure}

For each simulated dataset, we fit the standard method with $\beta_k=\beta_x$ for all $k$ (``Standard''), the parsimonious bivariate CAR (``Parametric'') model and the semi-parametric model with $\beta_k$ varying across $k$ using a B-spline basis expansion.  We compare two priors for $\sigma_b^2$, the variance of the coefficient process $\beta_k$.  The penalized complexity prior (``Semi -- PCP'') shrinks the process towards the constant function $\beta_k=\beta_x$ to avoid overfitting; the second prior for the variance induces a Uniform(0,1) prior on the proportion of overall model variance explained by variation in $\beta_k$ (``Semi--$R^2$'') to balance all levels of spatial confounding.  The prior distributions for all models are given in Supplemental Section \ref{s:app:priors}.  We fit the semi-parametric models for all $L\in\{1,5,10,20,30,40\}$ and select the number of basis functions using DIC \citep{spiegelhalter2002bayesian}.  All methods are fit using MCMC with 25,000 iterations and the first 5,000 discarded as burn-in.

\begin{figure}
\caption{{\bf Performance in the spectral domain for the discrete simulation study}: Median (solid) and 95\% confidence interval (dashed) for the sampling distribution of $\beta_k$ by eigenvalue $d_k$ for the standard (red), semi-parametric model with PC prior (green) and parametric bivariate CAR model (blue) for data generated with dependence between exposure and confounder controlled by $\beta_{xz}$ and kernel bandwidth $\phi$.  The black lines are the true value $\beta_x=0.5$.}\label{f:beta_hat_sim}
\centering
\includegraphics[page=1,width=.3\textwidth]{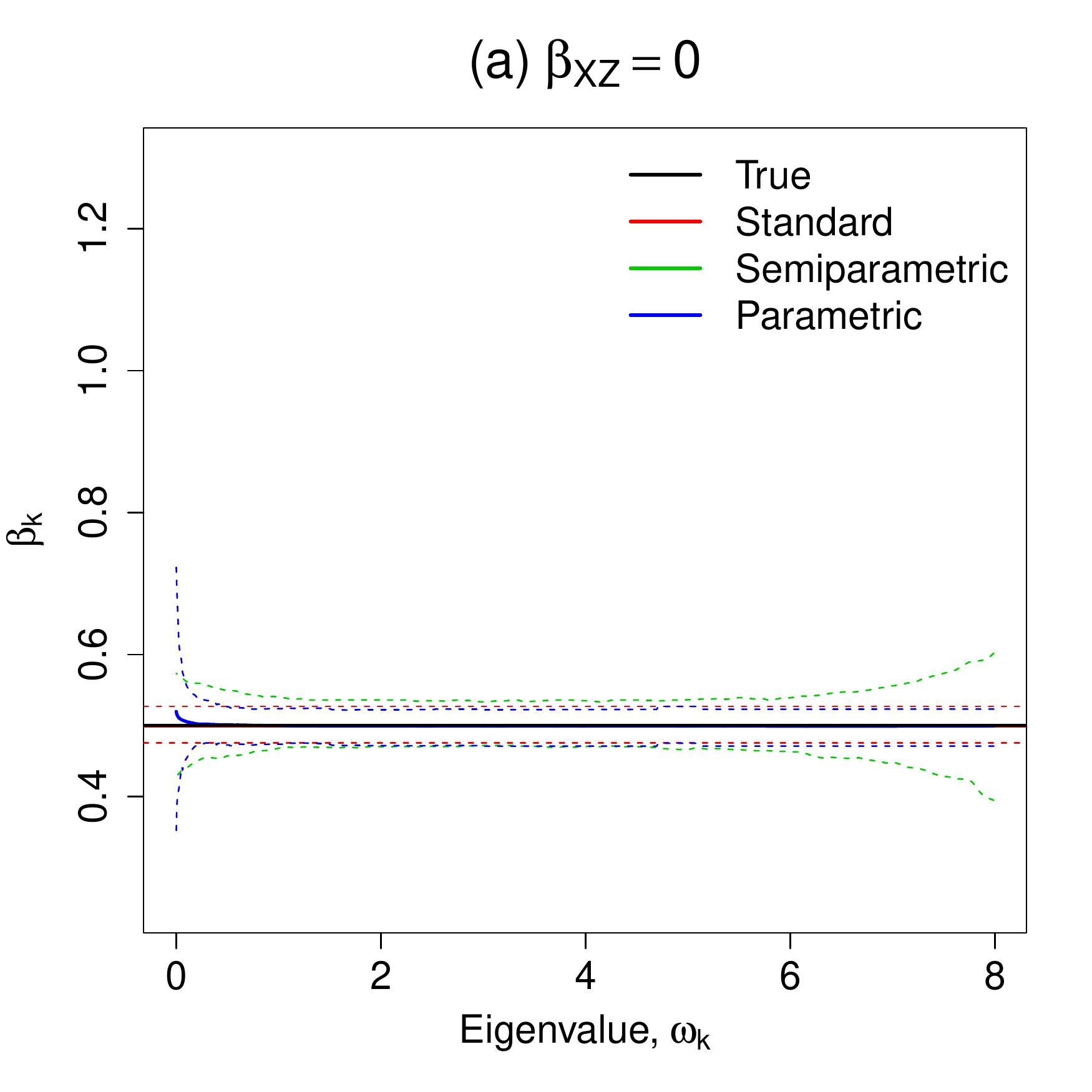}
\includegraphics[page=2,width=.3\textwidth]{figs/beta_hat_sim.pdf}
\includegraphics[page=3,width=.3\textwidth]{figs/beta_hat_sim.pdf}
\includegraphics[page=4,width=.3\textwidth]{figs/beta_hat_sim.pdf}
\includegraphics[page=5,width=.3\textwidth]{figs/beta_hat_sim.pdf}

\end{figure}

Table \ref{t:sim_disc} compares methods in terms of root mean squared error, bias, average (over datasets) posterior standard deviation and empirical coverage of 95\% intervals for $\beta_x$ and Figure \ref{f:beta_hat_sim} summarizes the sampling distribution of $\beta_k$ across $k$ for all methods and scenarios. The standard method performs well in the first scenario with no unmeasured confounder ($\beta_{xz}=0$), but in all other scenarios the standard method is biased and has coverage at or near zero.  The standard method allows for spatially-dependent residuals, but this does not eliminate spatial confounding bias.  Since the standard model assumes $\bX$ and $\bZ$ are independent and $\bX$ is included in the model, when $\bX$ and $\bZ$ are highly correlated all spatial variability is attributed to the treatment effect leading to bias and small posterior standard deviation.

The parametric model performs well in cases 1, 4 and 5 where there is no confounding at high frequencies.  In fact, the parametric model is nearly identical to the standard model in the first case with no spatial confounding, suggesting that little is lost by allowing for a parametric confounding adjustment when it is not needed.  However, the parametric model gives bias and low coverage in the cases with $\phi=1$ and thus the form of spatial confounding does not match the parametric CAR model.  The estimated $\beta_k$ curves in Figure \ref{f:beta_hat_sim} show that the parametric form of the $\beta_k$ model cannot match the slow decline in the true correlation of Figure \ref{f:corr_plots} when $\phi=1$.

\begin{table}\caption{{\bf Discrete-space simulation study results:} The standard method with constant effect across frequency and the proposed spectral method are evaluated in terms of their estimate of $\beta_x$ using root mean squared error (``RMSE''), bias (``Bias''), average posterior standard deviation (``SD'') and coverage of 95\% posterior intervals (``Cov'') for data generated with dependence between exposure and confounder controlled by $\beta_{xz}$ and kernel bandwidth $\phi$. Standard errors are in parentheses and all results are multiplied by 100.}\label{t:sim_disc}
\begin{center}\begin{tabular}{ll|ll|cccc}
Scenario & Method & $\phi$ & $\beta_{xz}$ & RMSE & Bias & SD & Coverage\\\hline 
1 & Standard      & -- &  0 & 1.3 ( 0.0) &  0.0 ( 0.1) &  1.3 ( 0.0) & 95.4 ( 0.9) \\ 
  & Parametric    &    &    & 1.3 ( 0.0) & -0.1 ( 0.1) &  1.3 ( 0.0) & 94.4 ( 1.0) \\
  & Semi - PCP    &    &    & 4.1 ( 0.3) &  0.0 ( 0.2) &  2.6 ( 0.1) & 95.2 ( 1.0) \\
  & Semi - $R^2$  &    &    & 4.0 ( 0.3) &  0.1 ( 0.2) &  2.4 ( 0.1) & 95.0 ( 1.0) \\
 \vspace{-6pt}&&&&&&\\
2 & Standard      & 1 & 1 & 19.0 ( 0.1) & 18.9 ( 0.1) &  1.4 ( 0.0) &  0.0 ( 0.0) \\
  & Parametric    &   &   & 16.1 ( 0.1) & 16.1 ( 0.1) &  1.5 ( 0.0) &  0.0 ( 0.0) \\
  & Semi - PCP    &   &   & 7.9 ( 0.2) & -0.7 ( 0.4) &  8.2 ( 0.0) & 95.4 ( 0.9) \\
  & Semi - $R^2$  &   &   & 8.0 ( 0.3) & -0.4 ( 0.4) &  8.8 ( 0.1) & 96.6 ( 0.8) \\
\vspace{-6pt}&&&&&&\\
3 & Standard      & 1 & 2 & 34.4 ( 0.1) & 34.4 ( 0.1) &  1.7 ( 0.0) &  0.0 ( 0.0)  \\
  & Parametric    &   &   & 27.1 ( 0.1) & 27.0 ( 0.1) &  1.9 ( 0.0) &  0.0 ( 0.0) \\
  & Semi - PCP    &   &   & 9.0 ( 0.3) &  0.7 ( 0.4) &  9.3 ( 0.0) & 96.4 ( 0.8) \\
  & Semi - $R^2$  &   &   & 9.3 ( 0.3) &  1.0 ( 0.4) &  9.4 ( 0.1) & 96.0 ( 0.9) \\
\vspace{-6pt}&&&&&&\\
4 & Standard      & 2 & 1 & 5.6 (0.1) & 5.4 (0.1) & 1.4 (0.0) & 4.0 (0.9) \\
  & Parametric    &   &   & 1.6 ( 0.0) &  0.4 ( 0.1) &  1.4 ( 0.0) & 90.4 ( 1.3) \\
  & Semi - PCP    &   &   & 8.4 ( 0.3) & -0.7 ( 0.4) &  9.3 ( 0.1) & 96.6 ( 0.8) \\
  & Semi - $R^2$  &   &   & 8.4 ( 0.3) & -0.2 ( 0.4) &  9.2 ( 0.1) & 95.8 ( 0.9) \\
\vspace{-6pt}&&&&&&\\
5 & Standard     & 2 & 2 & 8.8 (0.1) & 8.7 (0.1) & 1.5 (0.0) & 0.0 (0.0) \\
  & Parametric   &   &   & 1.5 ( 0.0) & -0.3 ( 0.1) &  1.5 ( 0.0) & 95.0 ( 1.0) \\
  & Semi - PCP   &   &   & 9.9 ( 0.3) & -0.8 ( 0.4) & 10.2 ( 0.1) & 94.2 ( 1.0) \\
  & Semi - $R^2$ &   &   & 9.9 ( 0.3) & -0.8 ( 0.4) & 10.3 ( 0.1) & 95.0 ( 1.0)\\
\end{tabular}\end{center}\end{table}

The semi-parametric methods have low bias and coverage near the nominal level for all five cases.  However, the posterior standard deviation is always larger for the semi-parametric models than the standard or parametric models.  Therefore, in these cases, the semi-parametric method is robust but conservative for estimating a casual effect in the presence of spatial confounding. Surprisingly, the semi-parametric methods are insensitive to the choice of prior.  Despite the PCP and $R^2$ prior having very different motivations, the results are similar under both priors, likely because $DIC$ often selects a small number of a basis functions which negates the influence of the prior for $\sigma_b^2$.

\subsection{Continuous-space}\label{s:sim_cont}

Data in the continuous space are generated similarly to the discrete case in (\ref{e:sim:data:disc}).  The data are simulated for $n=529$ spatial locations on a square $23 \times 23$ grid covering the unit square as
\begin{eqnarray}\label{e:sim:data:cont}
\bX &\sim& \text{Normal}\left({\bf 0},\sigma_x^2\Sigma_x\right),\\  
\bZ|\bX &\sim &\text{Normal}\left(\beta_{xz}\bW\bX,\sigma_z^2\Sigma_{z}\right),\nonumber\\  \bY|\bX,\bZ&\sim&\text{Normal}\left(\beta_x\bX+\beta_z\bZ,\sigma^2{\bf I}_n\right),\nonumber
\end{eqnarray}
where $\Sigma_j$ is the $n\times n$  $\Matern$ correlation matrix defined by parameters $\phi_j$ and $\nu_j$, and $\bW$ is the kernel smoothing matrix with bandwidth $\phi$ as in (\ref{e:sim:data:disc}).
In all cases we take $\sigma_x^2=\sigma_z^2=1$, spatial range parameters $\phi_x=\phi_{z}=0.1$, $\nu_x=\nu_{z}=0.5$, $\beta_x=\beta_{z}=1$ and $\sigma^2=0.25^2$, and we vary $\beta_{xz}\in\{0,1,2\}$, and the kernel bandwidth $\phi\in\{1/15,2/15\}$. For each combination of these parameters we generate 100 datasets. For each simulated dataset we fit four models: The standard $\Matern$ model (``Standard'') in Section \ref{s:cont:matern} with $\rho=0$ and thus no confounding adjustment, the bivariate $\Matern$ model (``Flexible $\Matern$'') with common range in (\ref{e:alpha_matern:common_range}), the parsimonious $\Matern$ model (``Parsimonious $\Matern$'') in (\ref{e:alpha_matern:parsi}) and the Gaussian mixture model (``Semiparametric'') in Section \ref{s:cont:semi}.  Prior distributions and computing details are given the Supplementary Sections \ref{s:app:priors} and \ref{s:app:comp}, respectively.

Table \ref{t:sim_cont} compares methods using the same metrics as in Section \ref{s:sim_disrete}.  The results mirror those in the discrete case.  The semi-parametric method maintains nearly the nominal coverage and low bias across all scenarios.  The parametric $\Matern$ models have bias and low coverage for the simulation settings where the data are not simulated with a $\Matern$ covariance. The flexible $\Matern$ dramatically reduces RMSE and improves coverage compared to the parsimonious model, but neither is sufficiently flexible for these cases.

\begin{table}\caption{{\bf Continuous-space simulation study results:} All methods are evaluated in terms of their estimate of $\beta_x$ using root mean squared error (``RMSE''), bias (``Bias''), average posterior standard deviation (``SD'') and coverage of 95\% posterior intervals (``Cov'') for data generated with dependence between exposure and confounder controlled by $\beta_{xz}$ and kernel bandwidth $\phi$. All results are multiplied by 100.}\label{t:sim_cont}
\begin{center}\begin{tabular}{ll|ll|cccc}
Scenario & Method & $\phi$ & $\beta_{xz}$ & RMSE & Bias & SD & Coverage\\\hline

1 & Standard  &  -  &  0  & 4.6 ( 0.4) &  0.1 ( 0.5) &  4.6 ( 0.0) & 96.0 ( 2.0) \\ 
& Flexible $\Matern$   &    &    & 14.3 ( 1.8) & -3.9 ( 1.4) & 13.0 ( 0.7) & 94.9 ( 2.2) \\ 
& Parsimonious $\Matern$  &    &    & 33.2 ( 3.0) & -6.5 ( 3.3) & 34.7 ( 1.2) & 90.9 ( 2.9) \\ 
& Semiparametric  &    &    & 6.9 ( 0.5) &  0.5 ( 0.7) &  6.9 ( 0.1) & 94.9 ( 2.2) \\    
 \vspace{-6pt}&&&&&&\\
2 & Standard     & 1/15 & 1 &   13.0 ( 0.5) & 11.8 ( 0.5) &  4.9 ( 0.0) & 35.0 ( 4.8) \\
  & Flexible $\Matern$    &    &    & 26.2 (  2.2) & -20.8 (  1.6) &  15.1 (  0.6) &  68.0 (  4.7) \\ 
  & Parsimonious $\Matern$   &    &    & 82.6 (  3.0) & -77.1 (  3.0) &  31.9 (  1.0) &   7.0 (  2.6) \\ 
  & Semiparametric  &    &    &  7.4 ( 0.6) &  0.3 ( 0.7) &  6.9 ( 0.1) & 93.0 ( 2.6) \\ 
 \vspace{-6pt}&&&&&&\\
3 & Standard     & 1/15& 2 & 17.9 ( 0.5) & 17.2 ( 0.5) &  5.2 ( 0.0) &  8.0 ( 2.7) \\ 
  & Flexible $\Matern$     &    &    & 50.6 (  2.6) & -45.2 (  2.3) &  21.4 (  0.8) &  27.0 (  4.5) \\ 
& Parsimonious $\Matern$   &    &    & 116.4 (   3.8) & -110.8 (   3.6) &   35.3 (   1.2) &    0.0 (   0.0) \\ 
& Semiparametric  &    &    &  6.7 ( 0.4) &  0.1 ( 0.7) &  6.9 ( 0.1) & 95.0 ( 2.2) \\ 
 \vspace{-6pt}&&&&&&\\
4 & Standard  &  2/15  &  1  &  5.9 ( 0.5) &  3.2 ( 0.5) &  4.7 ( 0.0) & 90.0 ( 3.0) \\ 
& Flexible $\Matern$   &    &    &  15.4 (  1.8) & -10.6 (  1.1) &  10.9 (  0.6) &  82.0 (  3.9) \\ 
& Parsimonious $\Matern$  &    &    &  49.2 (  3.5) & -34.8 (  3.5) &  33.3 (  1.0) &  76.0 (  4.3) \\ 
& Semiparametric  &    &    &  6.8 ( 0.4) &  0.4 ( 0.7) &  6.9 ( 0.1) & 95.0 ( 2.2) \\ 
 \vspace{-6pt}&&&&&&\\
5 & Standard  &  2/15  &  2  & 7.3 ( 0.4) &  4.7 ( 0.6) &  4.8 ( 0.0) & 77.0 ( 4.2) \\ 
& Flexible $\Matern$   &    &    & 16.3 (  1.1) & -12.2 (  1.1) &   9.4 (  0.4) &  70.0 (  4.6) \\ 
&  Parsimonious $\Matern$ &    &    & 63.0 (  3.8) & -54.7 (  3.1) &  28.8 (  1.2) &  30.0 (  4.6) \\ 
& Semiparametric  &    &    & 7.9 ( 0.6) & -0.8 ( 0.8) &  6.9 ( 0.1) & 90.0 ( 3.0) \\

\end{tabular}\end{center}\end{table}

\section{Real data examples}\label{s:examples}

We illustrate the methods by reanalyzing two publicly-available spatial datasets.  In Section \ref{s:lipcancer}, we analyze the well known lip cancer data from Scotland and we find different effects of the explanatory variable of interest, i.e. percentage of workforce in agriculture, fishing and forestry, for different spatial resolutions.  In Section \ref{s:covid}, we fit a model for COVID-19 mortality in the US and we find a consistent effect of air pollution exposure across resolutions.  All model fits in this section were carried out using the {\tt eCAR} package found in {\tt R} which  was created to fit the discrete space methods described in Sections \ref{s:discrete} and \ref{s:GLM}.

\subsection{Analysis of the Scottish lip cancer data}\label{s:lipcancer}
\begin{figure}
\caption{{\bf Lip cancer data}: Maps of Scotland district standard mortality ratio and percent of  work force in agriculture, fishing, and forestry (AFF).}\label{f:lipcancer}\vspace{6pt}
\hspace{0.35in}(a) Standardized mortality ratio 
\hspace{0.80in}(b) Percent of work force in AFF\\

\centering
\hspace{-1.1in}\includegraphics[page=1,width=.3\textwidth]{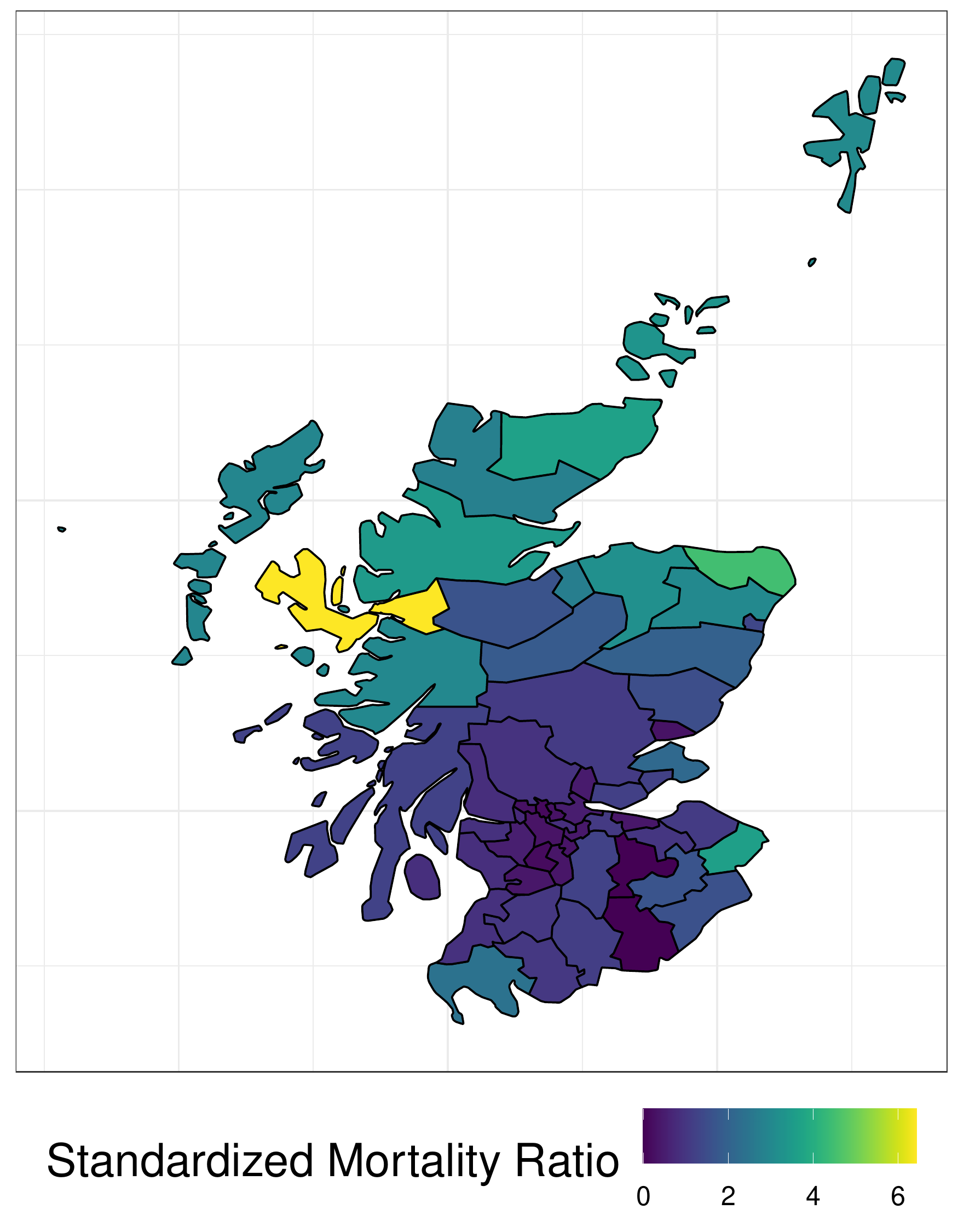}\hspace{.75in}
\includegraphics[page=1,width=.3\textwidth]{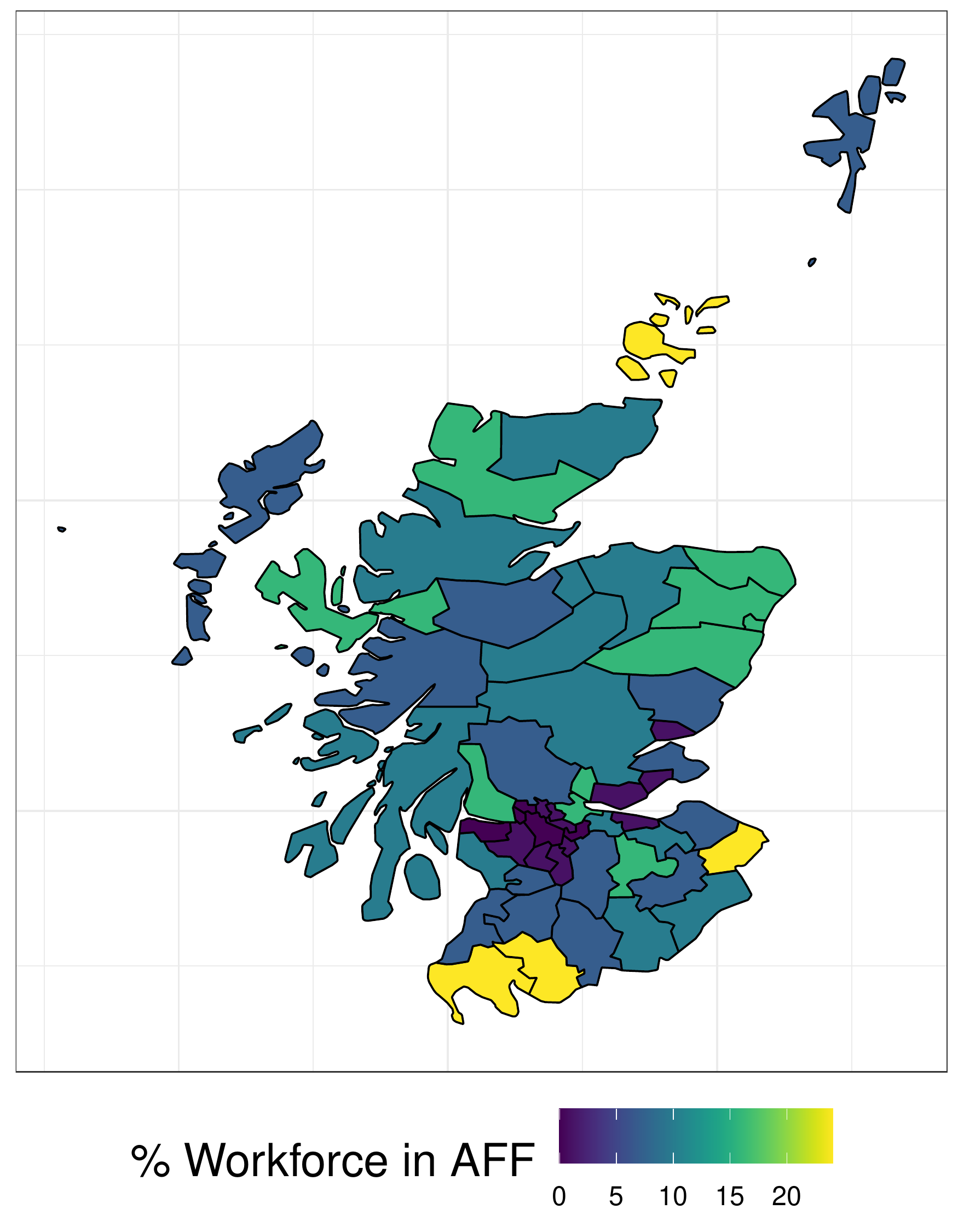}
\end{figure}

We first consider the well known lip cancer data.  These data are available in the {\tt CARBayesdata} package in {\tt R} and are displayed in Figure \ref{f:lipcancer}.  The data cover $n=56$ districts in Scotland. The three variables recorded for district $i$ are the number of recorded lip cancer cases, $Y_i$,  the  expected number of lip cancer cases computed using indirect standardisation based on Scotland-wide disease rates, $E_i$, and the percentage of the district's workforce employed in agriculture, fishing and forestry, $X_i$.  

We fit the spatial Poisson regression model ${Y}_i|\theta_i\indep\mbox{Poisson}\{E_i\exp(\theta_i)\}$, where $\theta_i$ is the log relative risk in district $i$.  As described in Section \ref{s:GLM} we model the log relative risks $\btheta=(\theta_1,...,\theta_n)^T$ as 
$$\btheta|\bX\sim\mbox{CAR} \{\beta_0{\bf 1}+\beta_x\bX + \bGamma\bA\bGamma^T\bX,\sigma_z^2,\lambda_z\},$$
where $\bA$ is diagonal with $k^{th}$ diagonal element $\alpha(\omega_k)$. 
Since the spatial domain is discrete, we fit the spectral models described in Section \ref{s:discrete} and studied in Section \ref{s:sim_disrete}.  The parametric method is fit using the same priors as those detailed in the simulation study of Section \ref{s:sim_disrete} and by collecting 20,000 MCMC iterates after discarding the first 5,000 as burn-in.  For the semi-parametric model we employ {\tt INLA} \citep{rue2009approximate} and use the PC prior with $L=10$ basis functions, chosen via DIC; results are stable for $L=\{10,20,30,40,50\}$.  We also fit two standard non-spectral methods where $\beta_k$ is constant over $\omega_k$, a Poisson regression with percent of workforce in AFF as covariate (``standard non-spatial'') and a Poisson regression that includes CAR random effects (``standard spatial'').

\begin{figure}
\caption{{\bf Effect of percent of workforce in AFF on lip cancer in Scotland}: Posterior mean (solid lines) and 95\% credible interval (dashed lines) of the effect $\exp(\beta_k)$ plotted against eigenvalue $\omega_k$ for spectral parametric, spectral semi-parametric (with $L=10$) and standard models; we consider two standard models, a Poisson regression on the percent of workforce in AFF (Standard non-spatial) and a Poisson regression with residuals modelled as CAR Leroux (Standard spatial).}\label{f:lipcancerbeta}
\centering
\includegraphics[page=1,width=.5\textwidth]{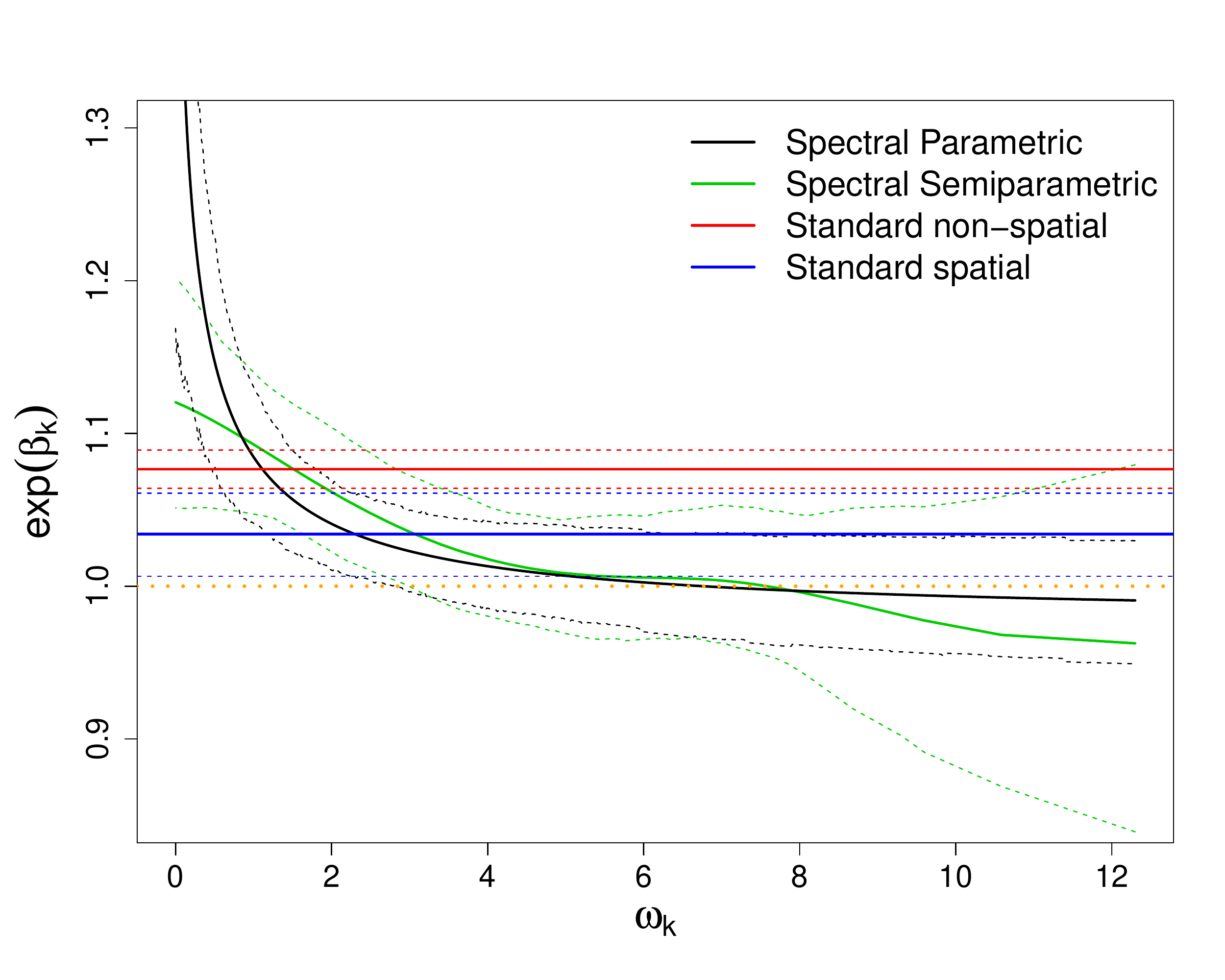}
\end{figure}

Figure \ref{f:lipcancerbeta} plots the posterior of $\exp(\beta_k)$ by the eigenvalue $\omega_k$ for each model. For the standard methods the posterior mean is positive and the 95\% interval excludes 1 indicating significant increase in risk for lip cancer for a unit increase in percent AFF. 
The spectral methods, which attempt to account for spatial confounding, do not agree with the standard methods: the estimated $\exp(\beta_k)$ trends toward 1 for large $\omega_k$ meaning that the results of the standard models should be interpreted with caution because the strength of the relationship between these variables is weak at the local (high-resolution) scale. These results are consistent with a missing confounding variable with the same large-scale spatial pattern as lip cancer disease and percentage of workforce in AFF. 


\subsection{Analysis of COVID-19 mortality and PM$_{2.5}$ exposure}\label{s:covid}

\cite{wu2020covidPM25} noticed that many of the pre-existing conditions that increase the mortality risk of COVID-19 are connected with long-term exposure to air pollution.  Thus, they conducted a study and found that an increase of 1 $\mu g/m^3$ in ambient fine particulate matter (PM$_{2.5}$) is associated with a 15\% increase in the COVID-19 mortality rate.  To further illustrate our proposed methods, we analyze the data collected by \cite{wu2020covidPM25} in an attempt to estimate the causal effect of PM$_{2.5}$ on COVID-19 mortality using spatial methods.  

\begin{figure}
\caption{{\bf PM$_{2.5}$ exposure and COVID-19 mortality by US county}: (a) Average PM$_{2.5}$ over 2000-2016 and (b) the log COVID-19 mortality rate (i.e., log(deaths/population)) through May 12, 2020 (counties with no deaths are shaded gray).}\label{f:covid}\vspace{12pt}
\centering
\hspace{-0in}(a) Average PM$_{2.5}$ ($\mu g/m^3$) \\
\includegraphics[scale=0.5,page=1]{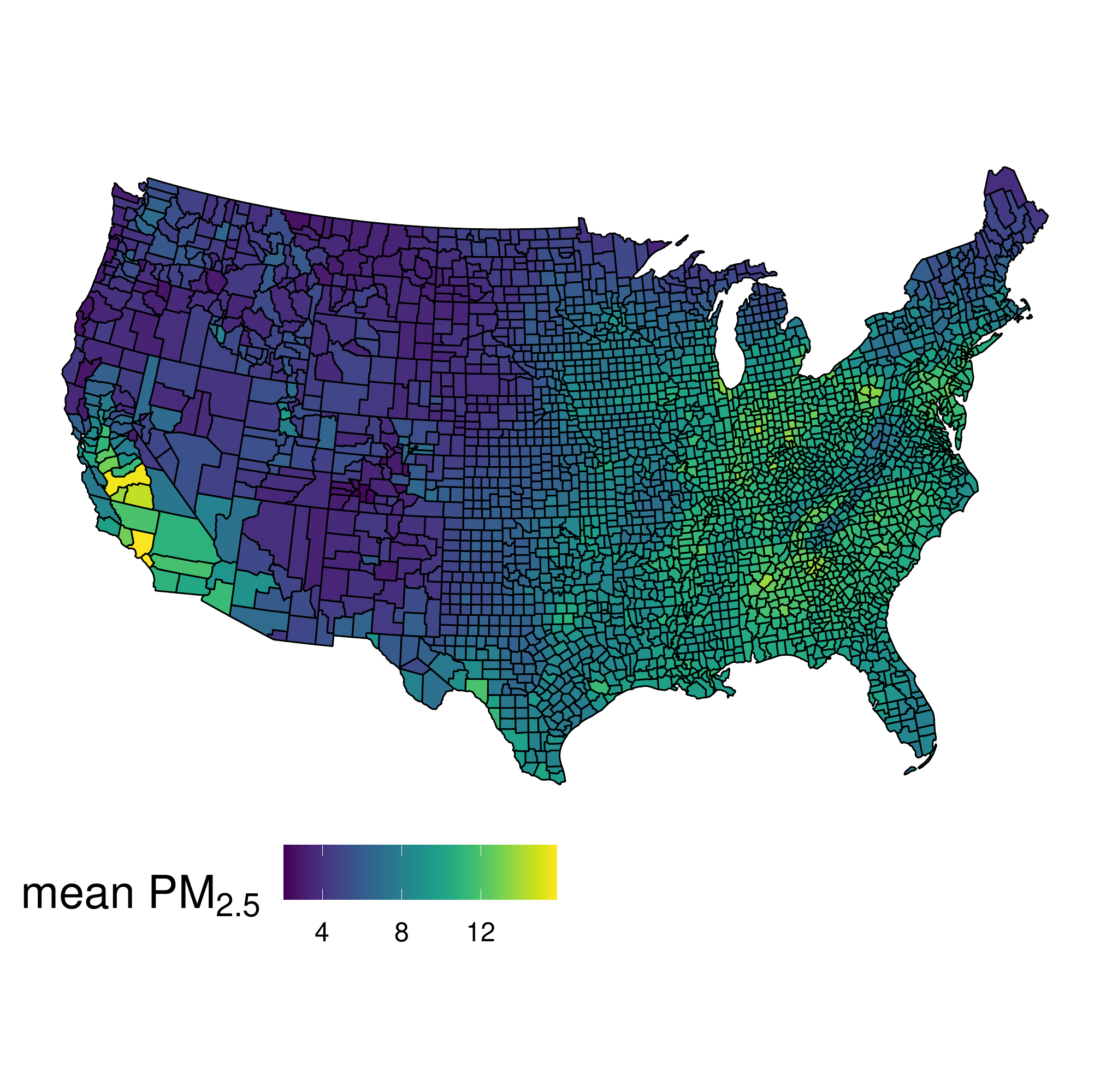}\\
(b) Log mortality rate\\
\includegraphics[scale=0.5,page=1]{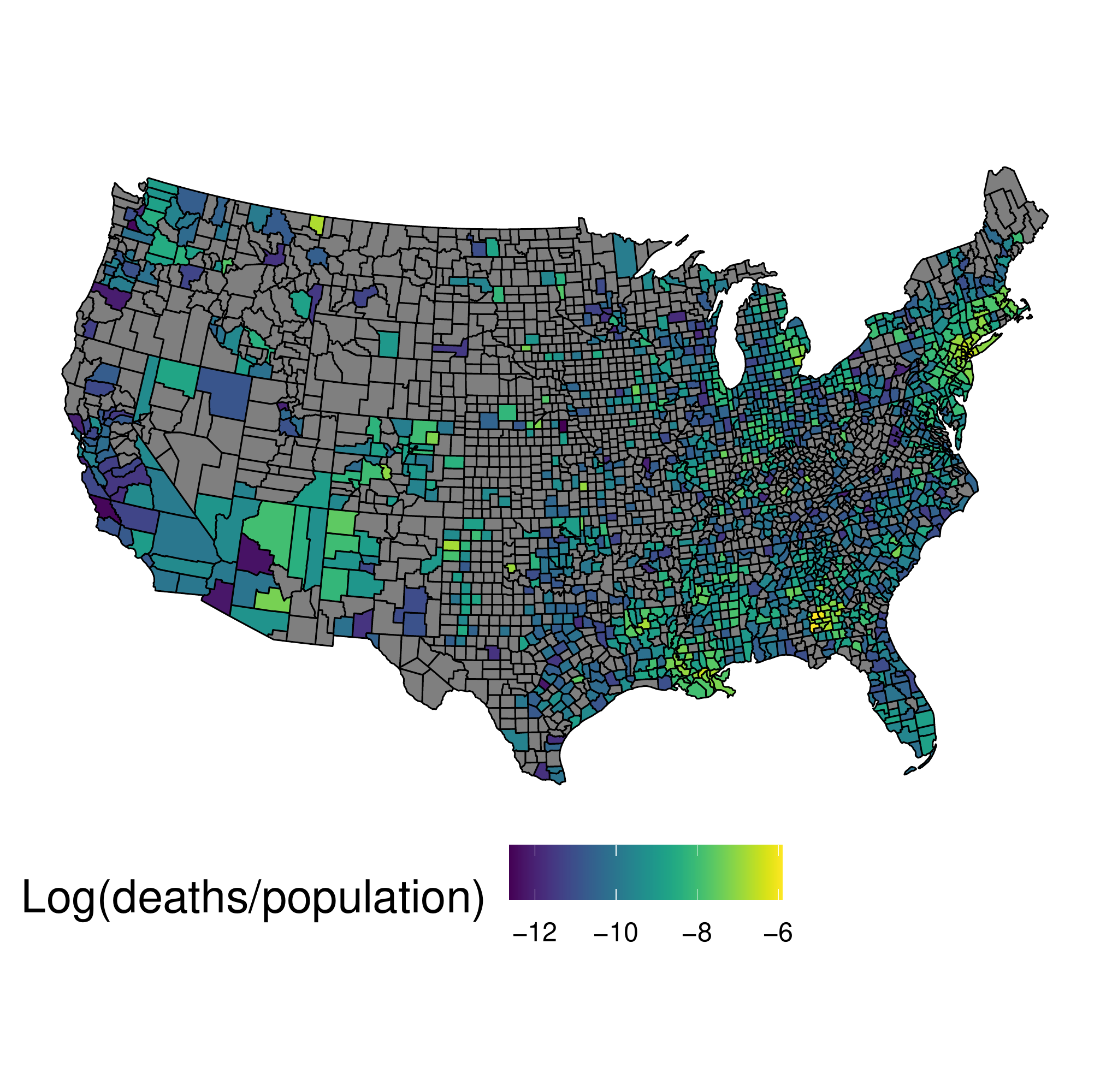}
\end{figure}

The response variable is the cumulative COVID-19 mortality counts through May 12, 2020 for US counties.   County-level exposure to PM$_{2.5}$ was calculated by averaging results from an established exposure prediction model for years 2000-2016 (see \citealt{wu2020covidPM25} for more details). Eight counties and twelve Virginia cities were missing from the database so we imputed their values using neighborhood means with neighbors defined by counties that share a boundary.  This resulted in mortality counts and PM$_{2.5}$ measures for $n= 3109$ counties (Figure \ref{f:covid}).  The long-term average  PM$_{2.5}$ is the highest in the Eastern US and California, while the mortality response is the highest in the New York, Los Angeles and Seattle areas.  The average PM$_{2.5}$ rate is a smoother spatial process than mortality, likely because the PM$_{2.5}$ exposure estimates are generated from predictive models.   In addition to PM$_{2.5}$ exposure, 20 additional potential confounding variables (e.g., the percentage of the population at least 65 years old) are included in our modeling (see \citealt{wu2020covidPM25} for the complete set of potential confounding covariates).

For county $i$, denote $Y_i$ as the number of deaths attributed to COVID-19, $E_i$ as the population, $X_i$ as the average PM$_{2.5}$ and $\bC_i$ as the vector of 20 known confounding variables.  Similar to \cite{wu2020covidPM25}, we fit a Negative-Binomial regression model $Y_i | X_i,Z_i,\bC_i \stackrel{indep}{\sim} \mbox{NegBin}\{r_i, p_i\}$, where $r_i$ is the size parameter (i.e., the number of successful trials) and $p_i$ the probability of success in each trial.  Under this model the mean is $\mbox{E}(Y_i|X_i,Z_i,\bC_i)= \lambda_i = r_i\frac{1-p_i}{p_i}$. We parameterize the model in terms of $\lambda_i$ and the over-dispersion parameter $r_i$.  The over-dispersion parameters have priors $\log(r_i) \sim N(0, 10)$ and the mean is linked to the linear predictor as $ \log(\lambda_i) = \log(E_i) + \theta_i$ where $\theta_i = \beta_x X_i + Z_i + \bC'_i\bm{\beta}_c$ and the offset term $E_i$ is the county population and $\bm{\beta}_c$ is a vector of regression coefficients associated with the confounding variables.  Following Section \ref{s:GLM}, the linear predictor $\bm{\theta}$ becomes 
\begin{equation}
\label{e:eta_covid}
\boldsymbol{\theta}|\bX  \sim \mbox{CAR}\left(\beta_0\bm{1} + \beta_x\bX + \bGamma\bA\bGamma^T\bX + \bbeta_c \bC, \sigma_z^2,\lambda_z \right), 
\end{equation}
where $\bC$ is a design matrix that includes an intercept term.   We fit this model using the  parametric and semi-parametric approaches detailed in Sections \ref{s:discrete:biCAR} and \ref{s:discrete:semi}.    We briefly comment that the Negative-Binomial modeling choice was motivated by our desire to mimic, as much as possible, the analysis found in \cite{wu2020covidPM25}.  However, we also considered a Binomial and Poisson model and inferences were relatively unchanged.

For the parametric model the prior distributions used in the simulation are employed.  It took approximately 3 hours to collect 20,000 MCMC iterates after discarding the first 5,000 as burn-in.    For the semi-parametric model we adopt the PC prior method described in Section \ref{s:sim_disrete} and use INLA \citep{rue2009approximate} to fit the model, which took approximately 5 minutes to run. We also fit a variant of the model employed in \cite{wu2020covidPM25} that we refer to as the ``standard model'', i.e., a Negative-Binomial regression with all control variables and a county random effects modeled using the CAR model.  Following \cite{wu2020covidPM25}, two separate analysis using all $n=3109$ counties and $n=1977$ counties that reported at least 10 confirmed COVID-19 deaths were conducted; this was done to account for the fact that the size of an outbreak in a given county may be positively associated both to COVID-19 mortality rate and to PM$2.5$, thus introducing confounding bias. 

\begin{figure}
\caption{{\bf Results from the parametric and semi-parametric approaches applied to the COVID-19 death counts reported up through May 12, 2020}: The posterior mean and 95\% credible interval of the mortality rate ratio associated to an increase of $1 \mu g/ m^3$ of PM$_{2.5}$, $\exp(\beta_{k})$, are plotted against $\omega_k$ for the semi-parametric fitted in INLA and the parametric model using MCMC. Column (a) are fits based on all $n=3109$ counties and column (b) are fits based on the $n=1977$ counties that reported at least 10 confirmed COVID-19 deaths. The fit from the standard approach refers to a regression model including all confounders and spatially structured random effects following a CAR model.}\label{f:covid_beta}
\centering
\includegraphics[page=1,width=.48\textwidth]{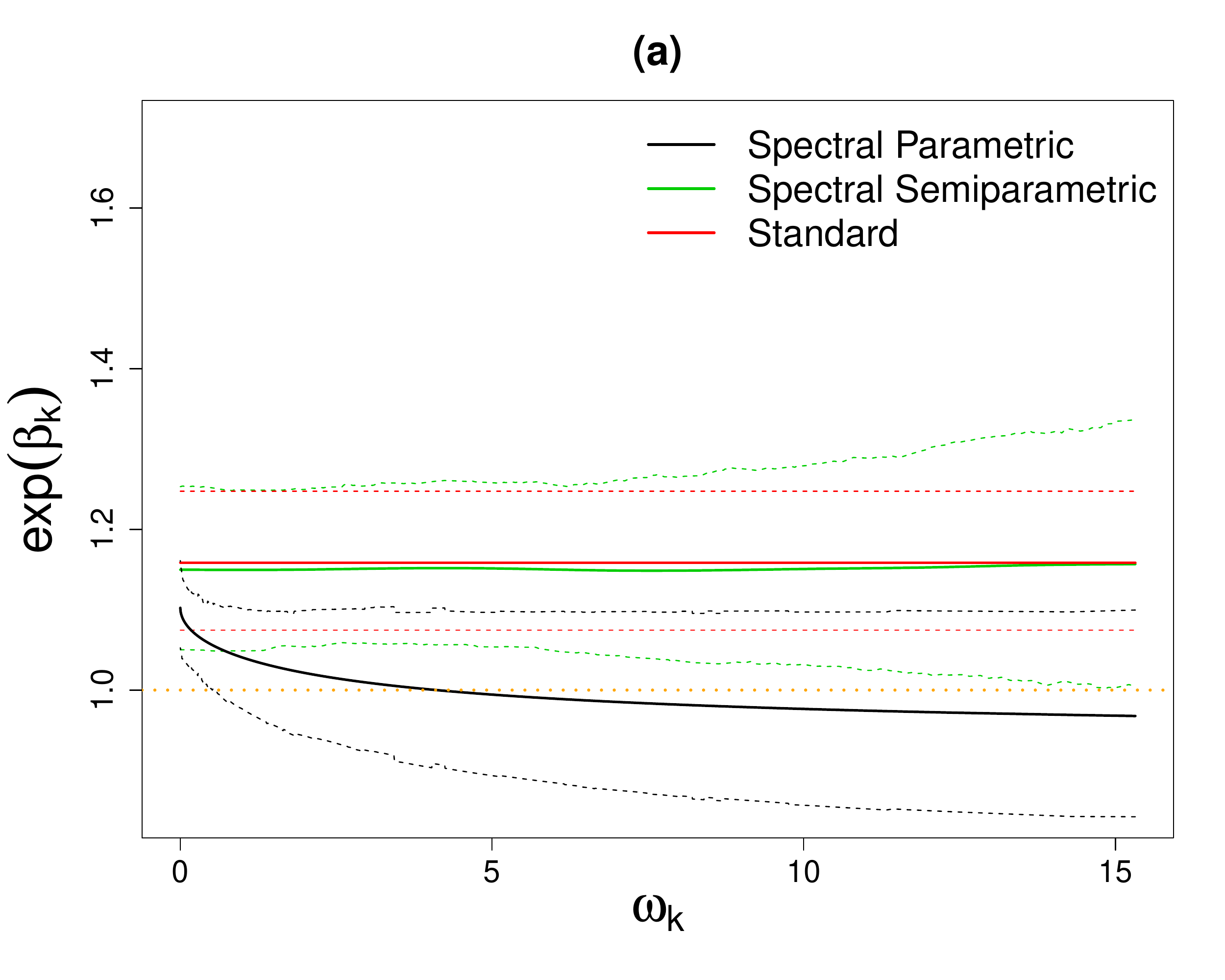} 
\includegraphics[page=1,width=.48\textwidth]{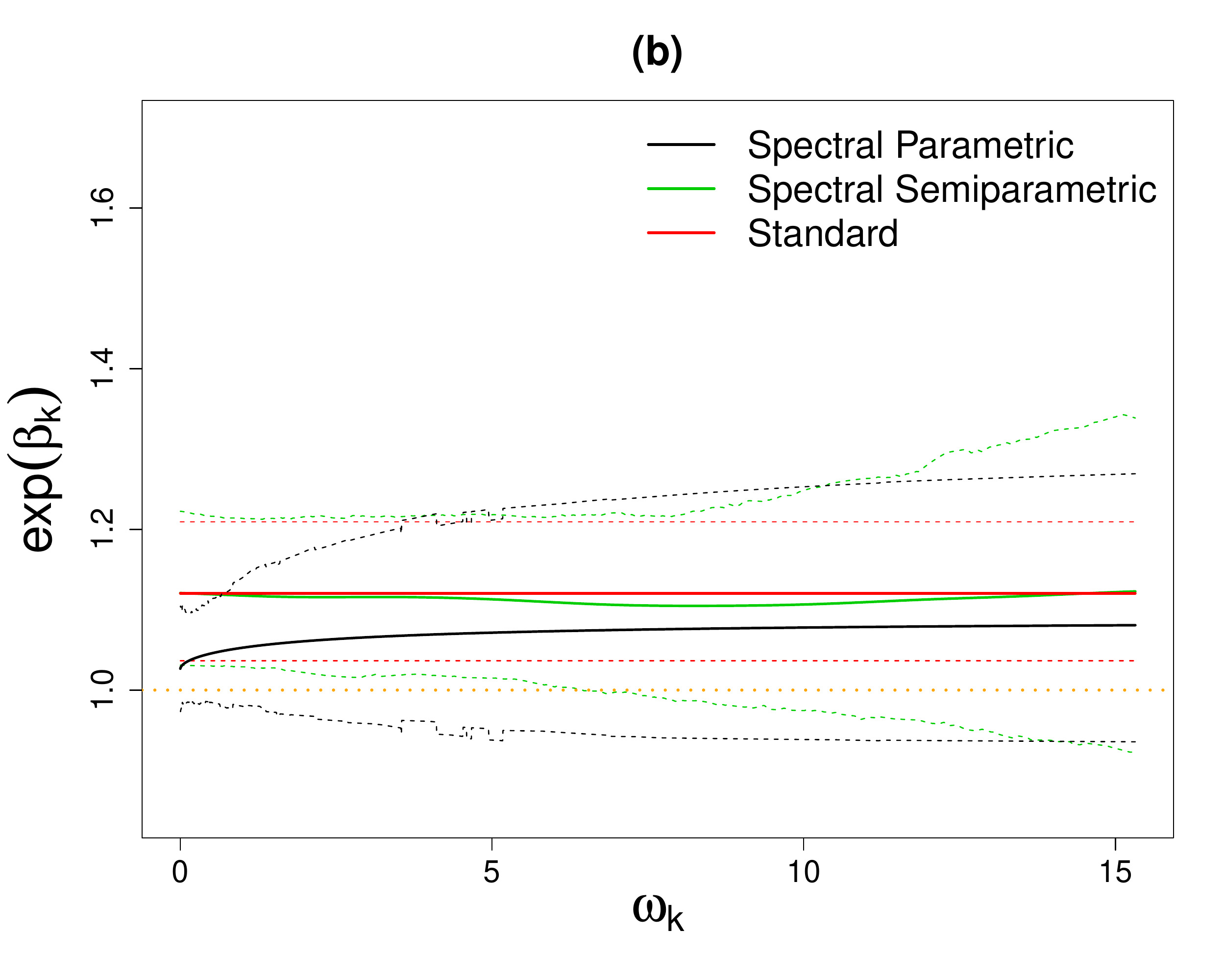}
\end{figure}

Figure \ref{f:covid_beta}a displays model fits using the full dataset while Figure \ref{f:covid_beta}b displays results when using the reduced data. The estimated increase in COVID-19 mortality rate, associated to an increase of $1 \mu g/ m^3$ of PM$_{2.5}$, under the standard model is $13\%$ (95$\%$ CI: $1.05, 1.21$) and $11\%$ (95$\%$ CI: $1.03, 1.19$) in the full and reduced analyses, respectively.  The posterior mean estimates from the parametric and semi-parametric spectral models generally agree with the standard approach, but the posterior standard deviation is higher for the spectral methods. 
In this analysis, the spectral methods support the standard spatial model and serve as a check of sensitivity to adjustments for missing confounding variables.

\section{Discussion}\label{s:disc}

In this paper, we propose new spectral methods to adjust for unmeasured spatial confounding variables.  We model the coherence of the treatment and unmeasured confounding variables as a function of the spatial resolution and introduce sufficient conditions that ensure the causal effect is identifiable.  These ideas are developed for continuous and discrete spatial domains, parametric and semi-parametric frameworks, and Gaussian and non-Gaussian data.  Simulation studies confirm that the proposed methods lower bias and provide valid inference if the model assumptions approximately hold, and two applications illustrate the use of the methods in practice.   

We compare parametric and semi-parametric approaches throughout the paper, but we recommend gravitating towards the semi-parametric approach.  The assumptions that permit estimating the causal effect differ for the parametric and semi-parametric approaches.   The parametric/parsimonious model depends on scale-invariant coherence between the regression coefficient and the unmeasured confounder, while the semi-parametric model depends on the assumption that their coherence tends to zero for large frequencies. While neither of these assumptions are empirically verifiable, we believe the later assumption is easier to understand and justify in practice.  In addition to this conceptual difference, the semi-parametric methods are easier to implement computationally.  Because the semi-parametric confounding adjustment takes the form of a linear combination of constructed covariates, it is straightforward to pass these constructed variables into standard spatial computing packages at a trivial computational cost.

\section*{Acknowledgments}

This work was partially supported by the National Institutes of Health (R01ES031651-01,R01ES027892-01) and King Abdullah University of Science and Technology (3800.2).  

\bibliographystyle{rss}
\bibliography{spatial_causal} 

\newpage
\setcounter{section}{0}
\setcounter{figure}{0}
\setcounter{page}{1}

\begin{center}
{\huge A spectral adjustment for spatial confounding}\\\vspace{6pt}
{\Large Supplementary materials}
\end{center}

\section{Extension to multiple predictors}\label{s:app:multivariate}

\subsection{Parsimonious continuous-space model}

In the spatial domain, let $X_0(\bs)=Z(\bs)$ be the unmeasured confounder and $X_1(\bs),...,X_p(\bs)$ be the $p$ observed covariates.  Given the confounder, the model is $Y(\bs) = \beta_0 + \sum_{j=0}^p\beta_jX_j(\bs) + \varepsilon(\bs)$.  For frequency $\bomega$, let $\calX_{0}(\bomega)=\calZ(\bomega)$ be the Fourier transform of the unmeasured confounder and $\calX_1(\bomega),...,\calX_p(\bomega)$ be the the Fourier transforms of the measured covariates.  In the spectral domain the joint model in (\ref{e:Covxz}) is extended to $\mbox{Cov}\{\calX_j(\bomega),\calX_k(\bomega)\} = \Omega_{jk}\sigma_j\sigma_kf_{jk}(\bomega)$ where $\Omega=\{\Omega_{jk}\}$ is a $(p+1)\times(p+1)$ correlation matrix, $\sigma_j>0$ are standard deviation parameters and $f_{jk}(\bomega)=f_{kj}(\bomega)$ are spectral densities.   The multivaritate extension of the parsimonious model in Section \ref{s:spec} sets $f_{jk}(\bomega)=\sqrt{f_{j}(\bomega)f_{k}(\bomega)}$.  This is clearly a valid model because the $(p+1)\times(p+1)$ covariance of $\calX_0(\omega),...,\calX_p(\bomega)$ is $\bF(\bomega)\Omega\bF(\bomega)$, where $\bF(\bomega)$ is diagonal with $j^{th}$ diagonal element equal $\sigma_j\sqrt{f_j(\bomega)}$, which is positive definite.

Blocking $\Omega$ to have first row column $(1,\Omega_{zx}^T)$ and bottom right $p\times p$ matrix $\Omega_x$ and marginalizing over $\calX_0(\bomega)=\calZ(\bomega)$ gives
\begin{equation}\label{e:ygivenx:pars:MV}
\calY(\bomega)|\calX_1(\bomega),...,\calX_p(\bomega) \indep \mbox{Normal}\left[\sum_{j=1}^p\{\beta_j +\alpha_j(\bomega)\}\calX_j(\bomega),q_0\sigma_z^2f_z(\bomega)\right],
 \end{equation}
where $\alpha_j(\bomega) = \frac{\sigma_z\sqrt{f_z(\bomega)}}{\sigma_j\sqrt{f_j(\bomega)}}q_j\calX_j(\bomega)$, $(q_1,...,q_p)=\Omega_{zx}^T\Omega_{x}^{-1}$ and $q_0 = \Omega_{zz}-\Omega_{zx}^T\Omega_{x}^{-1}\Omega_{zx}$.  In the spatial domain this the causal adjustment for covariate $j$ becomes ${\hat \bZ}_j=w_jC_{0j}C_{jj}^{-1}\bX_j$ where $C_{jk}=\mbox{Cov}(\bX_j,\bX_k)$ is defined by the $\Matern$ covariance parameters. The spatial model is 
$$Y(\bs_i)=\beta_0 + \sum_{j=1}^p\{X_j(\bs_i)\beta_j + {\hat Z}_j(\bs_i)\} + V(\bs_i) + \varepsilon(\bs_i)$$
where $V$ is a mean-zero Gaussian process with variance $\sigma_z^2q_o$ and $\Matern$ correlation with smoothness $\eta_z$ and range $\phi$ and $\varepsilon(\bs)\iid\mbox{Normal}(0,\sigma^2)$. 

\subsection{Semi-parametric continuous-space model}
Extending the semi-parametric model in Section \ref{s:cont:semi} to have multiple covariates is straightforward because the method is defined via the conditional distribution of the missing confounder given the covariates, rather than the joint distribution of the confounder and the covariates. We regress the confounder  onto the $p$ covariates using the additive model
\begin{equation*}
    \mbox{E}\{\calZ(\omega)\} = \sum_{j=1}^p\alpha_j(\omega)\calX_j(\omega) = \frac{1}{2} \sum_{j=1}^p\sum_{l=1}^L b_{jl}B_l(\omega)\calX_j(\omega).
\end{equation*}     
In this spatial domain this gives
$$Y(\bs) = \beta_0+\sum_{j=1}^p\beta_jX_j(\bs)+\sum_{j=1}^p\sum_{l=1}^L \beta_{jl} \hat{Z}_{jl}(\bs) + \delta(\bs),$$
where $\delta(\bs)$ is modeled as a Gaussian process and
$${\hat Z}_{jl}(\bs) = \sum_{\omega^f\in\mathcal{F}}  (2\pi\omega^f)^{\kappa+1} B_l(\omega^f)\bigtriangleup_{\mathcal{F}} \int  \frac{\mathcal{J}_\kappa(\omega^f ||\bs-\bs' ||)}{||\bs-\bs' ||^\kappa} X(\bs') d\bs'.$$
As in the univariate case, $\beta_j$ can be interpreted as the causal effect under the assumption that $\alpha_j(\bomega)\rightarrow 0$.

\subsection{Parsimonious CAR model}

In the discrete domain with multiple predictors, the oracle model is $\bY\sim\mbox{Normal}(\beta_0{\bf 1} +\sum_{j=1}^p\beta_j\bX_j + \bZ,\sigma^2{\bf I}_n)$.  Assuming marginal distribution $\bX_j\sim\mbox{CAR}(0,\sigma_j^2,\lambda_j)$ and $\bZ\sim\mbox{CAR}(0,\sigma_z^2,\lambda_z)$, parsimonious joint distributions in Section \ref{s:discrete:biCAR} and following the ideas above in the extension of the parsimonious $\Matern$ to the case of multiple predictors, the marginal distribution of $\bY$ over $\bZ$ is
$$\bY|\bX,\bV \sim \mbox{Normal}\left(\beta_0{\bf 1} + \sum_{j=1}^p\beta_j\bX_j + \sum_{j=1}^p\bGamma\bA_j\bGamma^T\bX_j + \bV,\sigma^2{\bf I}_n\right),$$
where $\bA_j$ is diagonal with $k^{th}$ diagonal element $\alpha_j(\omega_k) = w_j\frac{\sigma_z}{\sigma_j}\sqrt{\frac{1-\lambda_j+\lambda_j\omega_k}{1-\lambda_z+\lambda_z\omega_k}}$ and $\bV\sim\mbox{CAR}(0,q_0\sigma_z^2,\lambda_z)$.

\subsection{Semi-parametric CAR model}

Extending the semi-parametric model in Section \ref{s:discrete:semi} to the multivariate case follows the same steps as the semi-parametric model for continuous space above.   In the spectral domain with $\bZ^*=\bGamma\bZ$ and $\bX_j^*=\bGamma\bX_j$, a natural extension to the semi-parametric CAR model for the missing confounder is 
$$Z_k^*|X_{1k}^*,...X_{pk}^*\indep\mbox{Normal}\left(\sum_{j=1}^p\alpha_j(\omega_k) X_{j}^*,\frac{\sigma_Z^2}{1-\lambda_z+\lambda_z \omega_k}\right).$$ Defining $\beta_j(\omega) = \beta_{j}+\alpha_j(\omega) =     \sum_{l=1}^LB_{l}(\omega)b_{jl}$,
gives the spatial model
\begin{equation}\label{e:CAR:semi:Y2:MV}
 \bY|\bX,\bV\sim\mbox{Normal}
\left(\beta_0{\bf 1}+ \sum_{j=1}^p\sum_{l=1}^L{\hat \bZ}_{jl}b_{jl} + \bV,\sigma^2\bI_n\right).
\end{equation}
where $\bV\sim\mbox{CAR}({\bf 0},\sigma_z^2,\lambda_z)$.  Under the assumption that $\alpha_j(\omega_n)=0$, we summarize the causal effect of the $j^{th}$ covariate using the posterior of $\beta_{j}=\sum_{l=1}^LB_{l}(\omega_n)b_{jl}$.

\section{Spatial Causal framework}\label{s:app:PO}
Extending the notation in Section \ref{s:cont} of the main document and following the potential outcomes framework, we define  $Y_i(x_i)$ to be the potential outcome at location $\bs_i$ if the treatment at $\bs_i$ is $x_i$.  Similarly, define $\bY(\bx) = \{Y_1(x_1),\dots, Y_n(x_n)\}^T $ as the vector of potential outcomes at $\calS$ if treatment $\bx=(x_1,...,x_n)^T$ were received. For the potential outcomes to be well defined, we make the stable-unit-treatment-value assumption (SUTVA, Rubin (1980)), which states that there is no interference between units and there is a single version of each treatment level. In the spatial context, the no interference assumption implies that treatment applied at one location does not affect the outcome at other locations. 

Following the commonly used spatial regression model in the spatial statistics literature, we assume a linear additive structural model for the potential outcomes,
\begin{equation}\label{e:geo_obs_outcomes}
\bY(\bx) = \beta_0 + \beta_x \bx + \beta_z \bZ + \bvarepsilon, 
\end{equation}
where $\bvarepsilon=(\varepsilon_1,...,\varepsilon_n)^T$ and $\varepsilon_i\iid\text{Normal}(0,\sigma^2)$. The structural coefficient $\beta_x$ determines the causal relationship of the potential outcome and the treatment, and it is assumed to be spatially invariant. In this model, we do not specify the interaction of treatment and covariates, and therefore $\beta_x$ is both the marginal and conditional causal effect of treatment. Our objective is to estimate $\beta_x$.

To connect the observed and potential outcomes, we also make the consistency assumption that the observed outcomes are exactly the potential outcomes for the observed level of treatment $X_i$, $Y_i = Y_i(X_i)$. For the identifiability of $\beta_x$, we also make the conditional treatment ignorability assumption that $\bX \bot \bY(\bx) \mid \bZ$ for all $\bx$. The latter assumption holds if all factors that are associated with both the treatment and outcome variables are accounted for in $\bZ$. With these assumptions, the parameter $\beta_x$ in the regression model
\begin{equation}\label{e:obs_outcomes}
  \bY = \beta_0 + \beta_x\bX + \beta_z\bZ + \bvarepsilon.
\end{equation}
has a causal interpretation.

If we observe the confounder $\bZ$, then identification and estimation of $\beta_x$ is straightforward using multiple linear regression. However, we assume that $\bZ$ is an unmeasured confounder.   In the presence of an unmeasured confounder, $\beta_x$ is not identifiable in general.  Therefore, we propose to exploit the spatial structure of $\bZ$ to mitigate the effects of this unknown confounder.  
Although $\bZ$ confounds the causal relationship of $\bX$ and $\bY$ in the spatial domain, in Section \ref{s:spec} we propose assumptions in the spectral domain that identify the causal effect in the presence of an unmeasured spatial confounder.

\section{Identification under the parsimonious model}\label{s:app:IDparism}
The joint model  
\begin{eqnarray*}
\calY(\bomega)|\calX(\bomega) &\indep& \mbox{Normal}\left\{\left(\beta_x +\rho\frac{\sigma_z\sqrt{f_{z}(\bomega)}}{\sigma_x\sqrt{f_{x}(\bomega)}}\right)\calX(\bomega),(1-\rho^2
)\sigma_z^2f_z(\bomega)\right\}\\
\calX(\bomega) &\indep& \mbox{Normal}\left\{0,\sigma_x^2f_{x}(\bomega)\right\}
\end{eqnarray*}
is defined by parameters $\btheta=\{\beta_x,\rho,\sigma_x,\sigma_z,F_x,F_z\}$ where $F_j=\{f_j(\bomega); \bomega\in\calR^2\}$ and $\int f_j(\bomega)d\bomega=1$ for $j\in\{x,z\}$. Define $\mbox{E}\{\calY(\omega)|\calX(\bomega)\}=\mu_y(\bomega;\btheta)$, $\mbox{Var}\{\calY(\omega)|\calX(\bomega)\}=\tau^2_y(\bomega;\btheta)$
and $\mbox{Var}\{\calX(\omega)\}=\tau_x^2(\bomega;\btheta)$. 

Denote
$\btheta^{(j)} = \{\beta_x^{(j)},\rho^{(j)},\sigma_x^{(j)},\sigma_z^{(j)},F_x^{(j)},F_z^{(j)}\}$. The parameters $\btheta$ are identified if and only if $\mu_y(\bomega;\btheta^{(1)})=\mu_y(\bomega;\btheta^{(2)})$, $\tau_y^2(\bomega;\btheta^{(1)})=\tau_y^2(\bomega;\btheta^{(1)})$ and $\tau_x^2(\bomega;\btheta^{(1)})=\tau_x^2(\bomega;\btheta^{(1)})$ for all $\bomega$ implies that $\btheta^{(1)}=\btheta^{(2)}$.   First assume that $\tau_x^2(\bomega;\btheta^{(1)})=\tau_x^2(\bomega; \btheta^{(2)})$ for all $\bomega$.  Since $\int\tau_x^2(\bomega;\btheta)d\bomega=\sigma_x^2$, we must have $\sigma_x^{(1)}=\sigma_x^{(2)}$. This clearly implies that $F_x^{(1)}=F_x^{(2)}$, and thus the parameters in the marginal distribution of $\calX$ are identified.  Applying similar arguments to the $\tau_y^2(\bomega;\btheta)$ shows that $\tau_y^2(\bomega;\btheta^{(1)})=\tau_y^2(\bomega;\btheta^{(2)})$ for all $\bomega$ implies that $F_z^{(1)}=F_z^{(2)}$ and $\{1-(\rho^{(1)})^2\}(\sigma_z^{(1)})^2 = \{1-(\rho^{(2)})^2\}(\sigma_z^{(2)})^2$. 

The remaining parameters are identified by $\mu_y(\bomega;\btheta)$.  If we assume $\mu_y(\bomega;\btheta^{(1)})=\mu_y(\bomega;\btheta^{(2)})$ for all $\bomega$ and apply previous identifiability results that show $F_x^{(1)}=F_x^{(2)}$, $\sigma_x^{(1)}=\sigma_x^{(2)}$  and $F_z^{(1)}=F_z^{(2)}$, then the assumption that $f_x(\bomega)\ne f_z(\bomega)$ for some $\bomega$ implies that $\beta_x^{(1)}=\beta_x^{(2)}$ and $\rho^{(1)}\sigma_z^{(1)}=\rho^{(2)}\sigma_z^{(2)}$.  Combined with the result that $\{1-(\rho^{(1)})^2\}(\sigma_z^{(1)})^2 = \{1-(\rho^{(2)})^2\}(\sigma_z^{(2)})^2$, we have $\rho^{(1)}=\rho^{(2)}$ and $\sigma_z^{(1)}=\sigma_z^{(2)}$.

\section{Oracle confounder adjustment}\label{s:app:oracle}
We derive the oracle confounder adjustment for a given projection operator $\alpha(\bomega)$. If the inverse Fourier transform of $\alpha(\bomega)$ has a closed-form denoted by $K(\bs-\bs')$, then    \begin{eqnarray*}
    {\hat Z}(\bs) &=& \int_{\calR^2} \exp(i\bomega^T\bs){\hat \calZ}(\bomega)d\bomega \\
    &=&\int_{\calR^2} \exp(i\bomega^T\bs)\alpha(\bomega)\calX(\bomega)d\bomega\\
    &=& \int_{\calR^2} \exp(i\bomega^T\bs)\alpha(\bomega)\left[ \frac{1}{(2\pi)^d}\int_{\calR^2} \exp(-i\bomega^T \bs') X(\bs') d\bs' \right]d\bomega \\
     &=& \frac{1}{(2\pi)^d}\int_{\calR^2} \left[\int \exp(i\bomega^T(\bs-\bs'))\alpha(\bomega)d\bomega  \right]   X(\bs') d\bs'\\
     &=& \frac{1}{(2\pi)^d}\int_{\calR^2} K(\bs-\bs') X(\bs') d\bs'.
\end{eqnarray*} 

\section{Parameter identification for the bivariate $\Matern$ model}\label{s:app:IDmatern}
The parameters that define the marginal distribution of $X$, $\sigma_x$, $\nu_x$ and $\phi$, are identified following the arguments in Supplemental Section \ref{s:app:IDparism}.  Therefore, we assume they are fixed and known in this section. For the bivariate $\Matern$ model, define ${\mbox E}\{\calY(\bomega)| \calX(\bomega)\} = \mu(\bomega;\btheta)$ and ${\mbox V}\{\calY(\bomega)|\calX(\bomega)\} = \tau^2(\bomega;\btheta)$, where $\btheta = \{\beta_x,\rho,\nu_z,\nu_{xz},\sigma_z\}$ are the remaining unknown parameters.
 Defining $\delta = (1 + \phi^2||\bomega||^2)^{-1}$, the mean and variance are
\begin{eqnarray*}
     \mu(\bomega; \btheta) &=& \left(\beta_x + \rho\frac{\sigma_z\nu_{xz}}{\sigma_x\nu_x}\delta^{\nu_{xz}-\nu_x}\right) \calX(\bomega) \\
     \tau^2(\bomega; \btheta) & = & \sigma_z^2\nu_z\phi^2  \delta ^{\nu_z+1} \left[1-\rho^2\frac{\nu_{xz}^2}{\nu_x\nu_z} \delta ^{2\nu_{xz}-\nu_x-\nu_z}\right].
\end{eqnarray*}
Denote
$\btheta^{(j)} = \{\beta_x^{(j)},\rho^{(j)},\nu_z^{(j)},\nu_{xz}^{(j)},\sigma_z^{(j)}\}$.  The parameters $\btheta$ are identified if and only if $\mu(\omega;\btheta^{(1)})=\mu(\omega;\btheta^{(2)})$ and $\tau^2(\omega;\btheta^{(1)})=\tau^2(\omega;\btheta^{(1)})$ for all $\bomega$ implies that $\btheta^{(1)}=\btheta^{(2)}$. 

Now assume that $\mu(\bomega;\btheta^{(1)})=\mu(\bomega; \btheta^{(2)})$ and $\tau^2(\bomega;\btheta^{(1)})=\tau^2(\bomega; \btheta^{(2)})$ for all $\bomega$.  By assumption, $\nu_{xz}-\nu_x>0$, $2\nu_{xz}-\nu_x-\nu_z>0$ and $\rho^2\frac{\nu_{xz}^2}{\nu_x\nu_z}<1$.  Therefore,  $\mu(\bomega;\btheta)/\calX(\bomega) \rightarrow \beta_x$, and therefore $\beta_x^{(1)}=\beta_x^{(2)}=\beta_x^*$.  Considering $\mu(\bomega;\btheta_j)/\calX(\bomega)-\beta_x^*$ over $\bomega$ we have $\nu_{xz}^{(1)}=\nu_{xz}^{(2)}=\nu_{xz}^*$ and $\rho^{(1)}\sigma_z^{(1)}=\rho^{(2)}\sigma_z^{(2)}$. Turning to the variance terms, for large $\bomega$ we have $\tau^2(\bomega;\theta)\approx\sigma_z^2\nu_z\phi^2\delta^{\nu_z+1}$.  Therefore $\nu_z^{(1)}=\nu_z^{(2)}$ and $[\sigma_z^{(1)}]^2\nu_z^{(1)}=[\sigma_z^{(2)}]^2\nu_z^{(2)}$, and thus and $\sigma_Z^{(1)}=\sigma_Z^{(2)}$.  Finally, since $\rho^{(1)}\sigma_{Z}^{(1)}=\rho^{(2)}\sigma_{Z}^{(2)}$ from above we have $\rho^{(1)}=\rho^{(2)}$, and thus $\btheta_1=\btheta_2$.

\section{Parameter identification for the parsimonious CAR model}\label{s:app:IDparsiCAR}
We aim to establish the that parameters $\bm{\theta} = (\beta_x, \sigma^2, \rho, \lambda_x, \lambda_z, \sigma^2_x, \sigma^2_z)$ found in the joint model for $(\bY^*, \bX^*)$ that is partially described in \eqref{e:YparsiCAR}  are identifiable.  Here we will denote the joint model using $p(\bY^*, \bX^* | \bm{\theta})$.  As before, to establish identifiability we need to show that $p(\bY^*, \bX^* | \bm{\theta}^{(1)}) = p(\bY^*, \bX^* | \bm{\theta}^{(2)}) \Leftrightarrow \bm{\theta}^{(1)} = \bm{\theta}^{(2)}$. Note that we can express the joint model after setting $\beta_0=0$ w.l.o.g. in the following way
\begin{align*}
p(\bY^*, \bX^* | \bm{\theta}) & = p(\bY^*| \bX^*, \bm{\theta})p(\bX^*|\bm{\theta}) \\
& = N_n(\bY^*; \beta_x\bX^* + \rho\frac{\sigma_z}{\sigma_x}\bA\bX^*, \sigma^2_z(1-\rho^2)[(1-\lambda_z)\bm{I} + \lambda_z\bm{W}]^{-1} + \sigma^2\bm{I})\times \\ 
& \phantom{= \ \ } N_n(\bX^*; \bm{0},  \sigma^2_x[(1-\lambda_x)\bm{I} + \lambda_x\bm{W}]^{-1}).
\end{align*}
Here $N_n(\cdot;,\bm{m},\bm{V})$ denotes a $n$ dimensional multivariate density function with mean $\bm{m}$ and covariance matrix $\bm{V}$, and, as before, $\bA$ is a diagonal matrix whose $i$th diagonal entry is $\sqrt{\frac{1-\lambda_x + \lambda_x \omega_i}{1-\lambda_z + \lambda_z \omega_i}}$.  Note that we can establish identifiability by way of the mean structure, covariance structure, or both from the joint model.  Since $p(\bX^* | \bm{\theta})$ is simply a Leroux model and its parameters are known to be identifiable we have that $\sigma^2_x$ and $\lambda_x$ are both identifiable.  Thus, we focus on $(\beta_x, \sigma^2, \rho, \lambda_z, \sigma^2_z)$.  It is straightforward to see that
\begin{align*}
p(\bY^*| \bX^*, \beta_x, \sigma^{2(1)}, \rho, \lambda_x, \lambda_z, \sigma^2_x, \sigma^2_z) = p(\bY^*| \bX^*, \beta_x, \sigma^{2(2)}, \rho, \lambda_x, \lambda_z, \sigma^2_x, \sigma^2_z)\Leftrightarrow \sigma^{2(1)} = \sigma^{2(2)} 
\end{align*}
which implies that $\sigma^2$ identifiable.   

Next since matrix inverse is a bijection we can work with  $\sigma^{-2}_z(1-\rho^2)^{-1}[(1-\lambda_z)\bm{I} + \lambda_z\bm{W}]$ which can be expressed as $\sigma^{-2}_z(1-\rho^2)^{-1}\bm{I} + \sigma^{-2}_z(1-\rho^2)^{-1}\lambda_z(\bm{W} - \bm{I})$. Letting $\tau = \sigma^{-2}_z(1 - \rho^2)^{-1}$ and we have
\begin{align*}
   & \phantom{\Leftrightarrow \ \ } \tau^{(1)}\bm{I} + \tau^{(1)}\lambda^{(1)}_z(\bm{W} - \bm{I}) + \sigma^2\bm{I} - [\tau^{(2)}\bm{I} + \tau^{(2)}\lambda^{(2)}_z(\bm{W} - \bm{I}) + \sigma^2\bm{I}] = 0 \\
   & \Leftrightarrow  (\sigma^2 + \tau^{(1)} - (\sigma^2 + \tau^{(2)}))\bm{I} + (\tau^{(1)}\lambda^{(1)}_z - \tau^{(2)}\lambda^{(2)}_z)(\bm{W} - \bm{I}) = 0 \\
   & \Leftrightarrow  (\tau^{(1)} - \tau^{(2)}) + (\tau^{(1)}\lambda^{(1)}_z - \tau^{(2)}\lambda^{(2)}_z)(\omega_i - 1) = 0 \ \forall \ \omega_i \\
   & \Leftrightarrow  \tau^{(1)} - \tau^{(2)} = 0 \ \mbox{and} \  \tau^{(1)}\lambda^{(1)}_z - \tau^{(2)}\lambda^{(2)}_z = 0. 
\end{align*}
Now $(\tau^{(1)} - \tau^{(2)}) = 0$ establishes $\tau^{(1)} = \tau^{(2)}$ and $\tau^{(1)} = \tau^{(2)}$ along with $\tau^{(1)}\lambda^{(1)}_z - \tau^{(2)}\lambda^{(2)}_z=0$     establishes $\lambda^{(1)}_z = \lambda^{(2)}_z$.  Thus, $\tau$ and $\lambda_z$ are identifiable. In summary, to this point, we have shown that $\lambda_x$, $\lambda_z$, $\sigma^2_x$, $\sigma^2$ and $\tau$ are all identifiable.  We now show that $\beta_x$ and $\eta = \rho\sigma_z$ are identifiable.   Note that 
\begin{align*}
   & \beta_x\bX^* + \rho\frac{\sigma_z}{\sigma_x}\bA\bX^* = (\beta_x\bm{I} + \frac{\eta}{\sigma_x}\bA)\bX^*. 
\end{align*}
Thus, we can focus on $\beta_x\bm{I} + \frac{\eta}{\sigma_x}\bA$ such that
\begin{align*}
   & \phantom{\Leftrightarrow \ \ } (\beta^{(1)}_x\bm{I}  + \frac{\eta^{(1)}}{\sigma_x}\bA) -  (\beta^{(2)}_x\bm{I} + \frac{\eta^{(2)}}{\sigma_x}\bA) = 0  \\
   & \Leftrightarrow  (\beta^{(1)}_x - \beta^{(2)}_x)\bm{I}  + (\eta^{(1)} - \eta^{(2)})\frac{1}{\sigma_x}\bA = 0  \\
   & \Leftrightarrow  (\beta^{(1)}_x - \beta^{(2)}_x)  + (\eta^{(1)} - \eta^{(2)})\frac{1}{\sigma_x}\sqrt{\frac{1-\lambda_x + \lambda_x \omega_i}{1-\lambda_z + \lambda_z \omega_i}} = 0  \ \forall \ \omega_i \\
   & \Leftrightarrow  (\beta^{(1)}_x - \beta^{(2)}_x) = 0 \ \mbox{and} \ (\eta^{(1)} - \eta^{(2)}) = 0,
\end{align*}
establishing the identifiability of $\beta_x$ and $\eta=\rho\sigma_z$.  Now since $\tau = \sigma^{-2}_z(1-\rho^2)^{-1}$ and $\eta=\rho\sigma_z$, are identifiable, then so to are $\rho$ and $\sigma^2_z$ where  $\rho = \sqrt{\frac{\eta^2}{\tau^{-1} + \eta^2}}$ and $\sigma^2_z = \tau^{-1} + \eta^2$.

\section{Sample datasets for the simulation study}\label{s:app:datasets}

\begin{figure}
\caption{{\bf Simulated data in discrete space}: Realizations from the discrete-space simulation's data-generating process for different kernel bandwidth ($\phi$) and strength of exposure/confounder dependence ($\beta_{xz}$). }\label{f:sim_data}\vspace{12pt}
\centering
\hspace{-3.8in}(a) $\phi=1$ and $\beta_{xz}=1$  \\
\includegraphics[page=1,width=.27\textwidth]{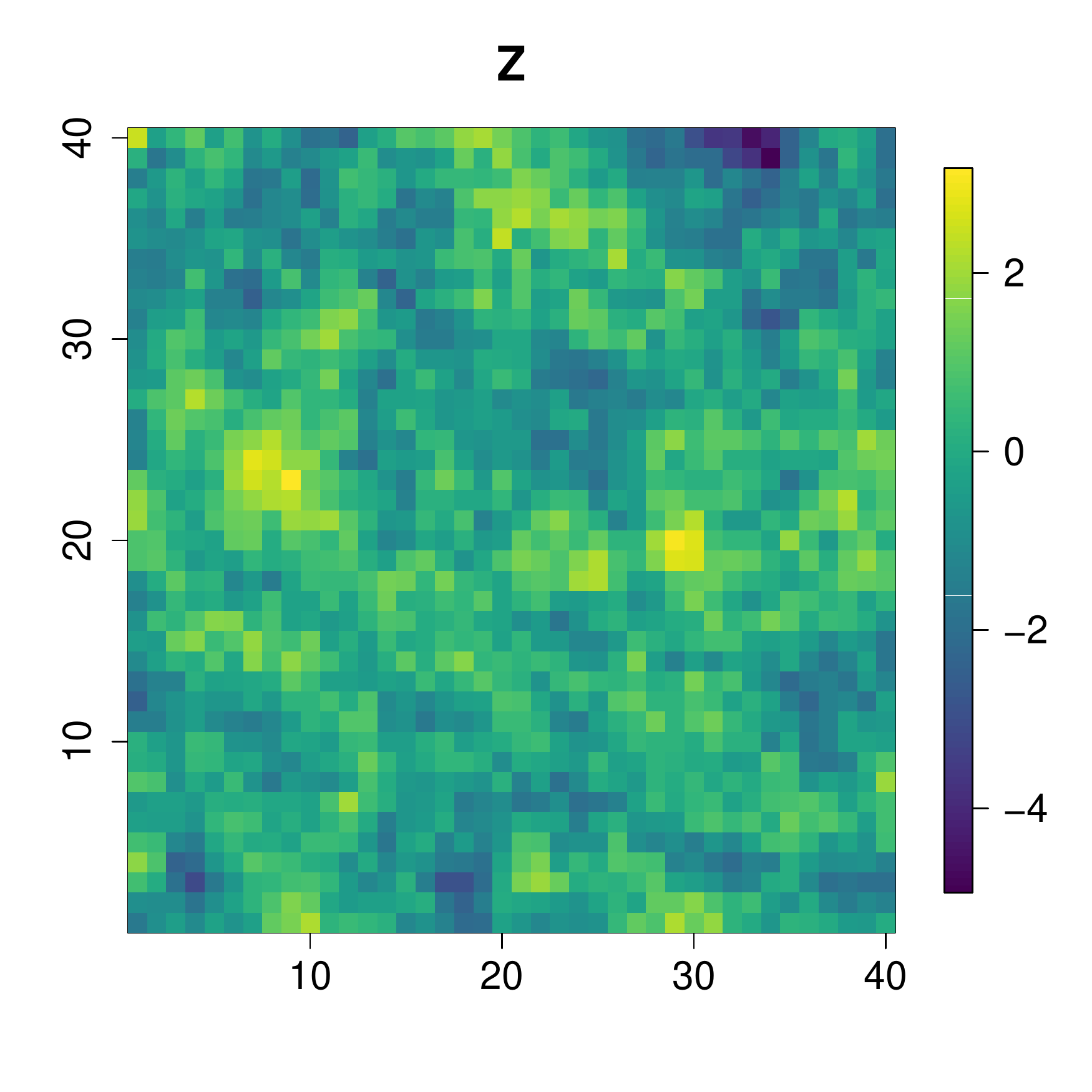}	
\includegraphics[page=2,width=.27\textwidth]{figs/sim2}	
\includegraphics[page=3,width=.27\textwidth]{figs/sim2}	
	
\hspace{-3.8in}(b) $\phi=1$ and $\beta_{xz}=2$ \\
\includegraphics[page=1,width=.27\textwidth]{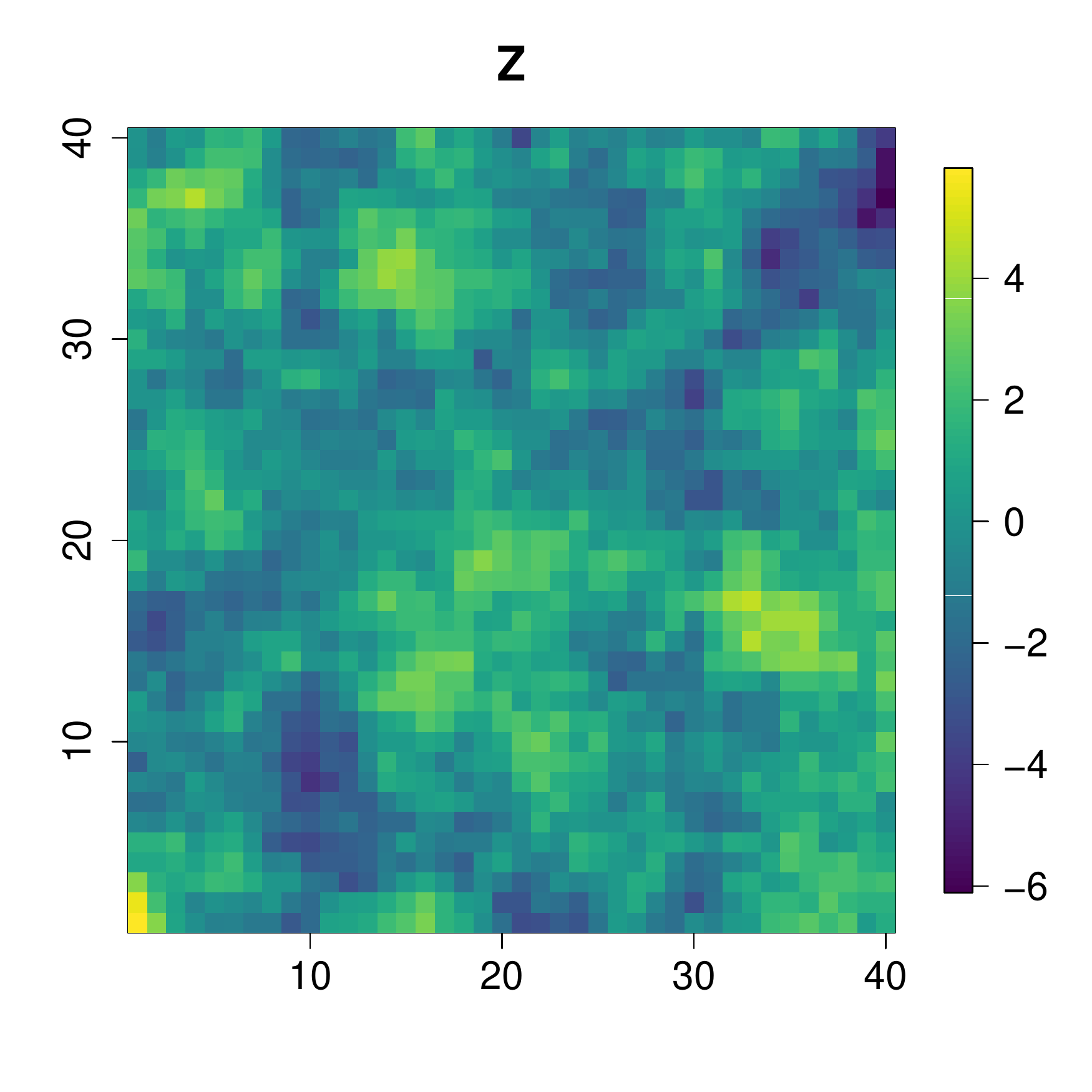}	
\includegraphics[page=2,width=.27\textwidth]{figs/sim3}	
\includegraphics[page=3,width=.27\textwidth]{figs/sim3}

\hspace{-3.8in}(c) $\phi=2$ and $\beta_{xz}=1$\\
\includegraphics[page=1,width=.27\textwidth]{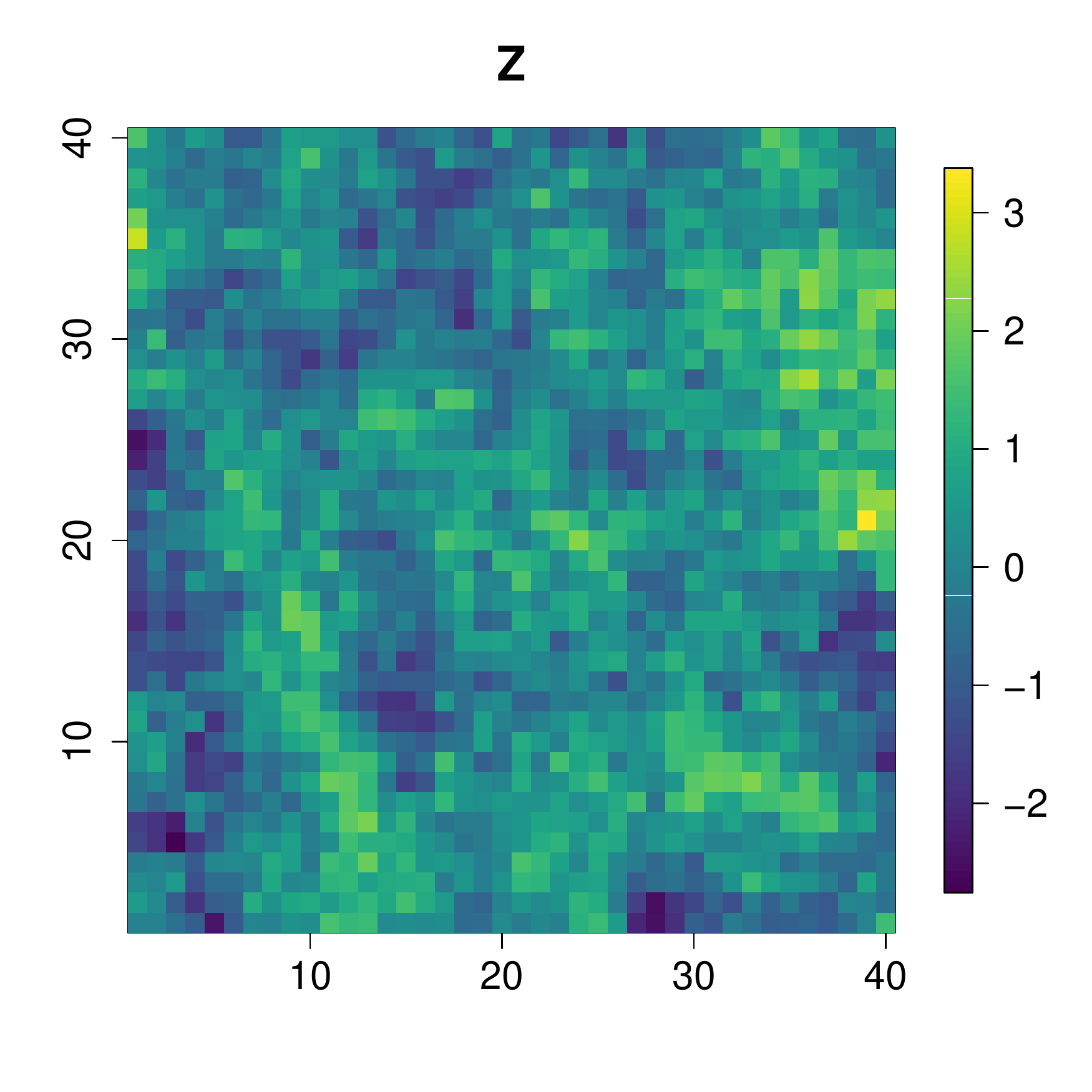}	
\includegraphics[page=2,width=.27\textwidth]{figs/sim4}	
\includegraphics[page=3,width=.27\textwidth]{figs/sim4}	

\hspace{-3.8in}(d) $\phi=2$ and $\beta_{xz}=2$\\
\includegraphics[page=1,width=.27\textwidth]{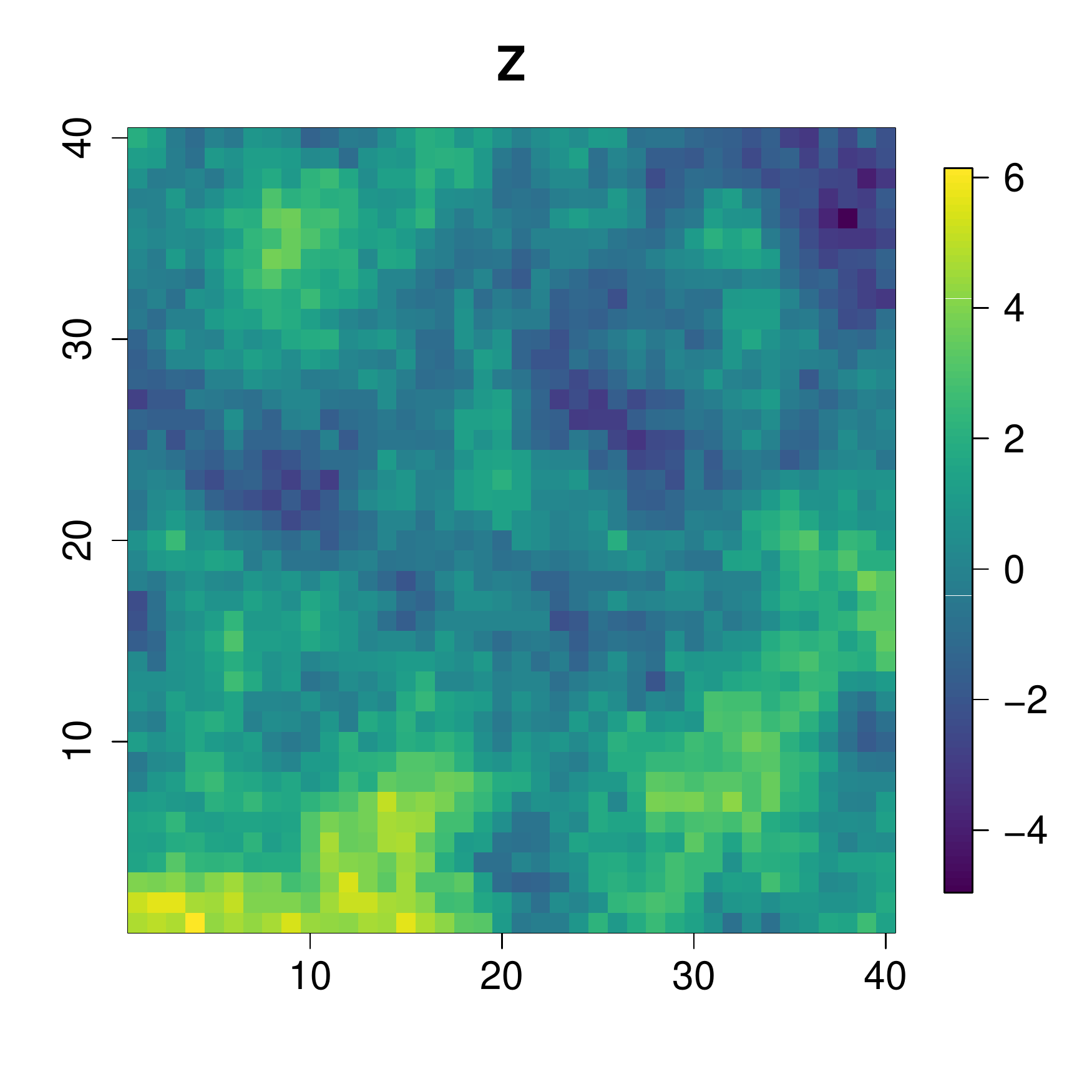}	
\includegraphics[page=2,width=.27\textwidth]{figs/sim5}	
\includegraphics[page=3,width=.27\textwidth]{figs/sim5}	
\end{figure}

Figure \ref{f:sim_data} plots a dataset from each simulated scenario in Section \ref{s:sim_disrete} to illustrate the spatial dependence of each variable and the relationships between variables.

\section{Prior distributions}\label{s:app:priors}

\subsection{Discrete cases}

For the Gaussian responses in the simulation study, the variances in the standard and parametric models are parameterized as $\sigma_z^2=\tau^2r$ and $\sigma^2=\tau^2(1-r)$ and the priors are $\beta_0,\beta_x\sim\mbox{Normal}(0,100)$, $\tau^2\sim\text{InvGamma}(0.1,0.1)$ and $r,\lambda_z\sim\mbox{Uniform}(0,1)$. For the parsimonious  bivariate CAR model we fix $\sigma_x=1$ and select priors $\lambda_x\sim\mbox{Uniform}(0,\lambda_z)$ and $\rho\sim\mbox{Uniform}(-1,1)$.  

In the parsimonious bivariate CAR model we used the following priors for the variance components  $\sigma^2 \sim \mbox{Gamma}(1, 1)$, $\sigma^2_x \sim \mbox{InvGamma}(1, 1)$, and $\tau = \sigma^2_z(1-\rho^2) \sim \mbox{Gamma}(1, 1)$ where both the Gamma and Inverse Gamma distributions are parameterized using shape and rate.  For spatial dependence parameters we used $\lambda_z \sim \mbox{Uniform}(0, 1)$ and $\lambda_x \sim \mbox{Uniform}(0, \lambda_z)$ (although we also ran simulation with $\lambda_x \sim \mbox{Uniform}(0, 1)$ and results where similar).  Finally for the regression type parameters we use $\beta_x \sim N(0,100)$ and  $\psi=\rho\frac{\sigma_z}{\sigma_x} \sim N(0, 100)$.  To improve mixing $(\beta, \psi)$ were updated in a blocked Gibbs step.  Apart from $\sigma^2_x$, all other parameters were updated with a random walk Metropolis step using a Gaussian distribution to generate candidate values. 

For the semi-parametric models with $L>1$ we use the CAR prior for $\bb$ given in Section \ref{s:cont:semi} with either a penalized complexity prior \citep[PCP;][]{simpson2017pcprior} or $R^2$ prior \citep{zhang2020bayesian} for $\sigma_b^2$, as described below. 
The semi-parametric CAR model in Section \ref{s:discrete:semi} is $\bY|\bV  \sim\mbox{Normal}(\beta_0+\beta_x\bX + {\tilde Z}\bb + \bV, \sigma^2\bI_n)$.  To set priors, we parameterize the covariance parameters as $\sigma_z^2=\tau^2c(\lambda_z)r$ and $\sigma^2=\tau^2(1-r)$ for $c(\lambda_z) = n/\sum_{k=1}^nf_{z}(\omega_k)$ and $r\in(0,1)$.  Under this parameterization, the total (over space) error variance is $$\mbox{Trace}\{\mbox{Cov}(\bV)+\sigma^2\bI_n\} = \tau^2nr+\tau^2n(1-r)=\tau^2n,$$
and thus $\tau^2$ controls the overall variance and $r$ is the proportion of variance attributed to the spatial component of the error. Also, assume the prior for the adjustment coefficients $\bb$ is normal with mean zero and precision $\sigma^{-2}_b\Omega$, where $\Omega$ has diagonal elements $N_l$ and $(j,l)$ off-diagonal element $-I(|j-l|=1)$.  
The first is the penalized complexity prior \citep[PCP;][]{simpson2017pcprior} on the standard deviation $\sigma_b$, i.e. $\sigma_b\sim\mbox{Exponential}(\xi)$ with scale parameter $\xi=-\log(0.01)0.31/U$, with $U=0.5$; this choice sets the marginal standard deviation for $\bb$ approximately equal to $0.5$, following the rule of thumb proposed by \cite{simpson2017pcprior}. The PCP shrinks the model to the simpler base model without a confounder adjustment. 
The second is based on the R2D2 prior of \cite{zhang2020bayesian}.  
The variance is written $\sigma_b^2=\tau^2\sigma_R^2$ where the  prior density for $\sigma_R^2$ is $f(\sigma_R^2) \propto (\sigma_R^2+1)^{-2}$, which induces a Uniform(0,1) prior on the proportion of variance explained by the confounding adjustment,
$\mbox{Trace}\{\mbox{Cov}({\tilde \bZ}\bb)\}/\mbox{Trace}\{\mbox{Cov}({\tilde \bZ}\bb)+\mbox{Cov}(\bV) + \sigma^2\bI\} = \sigma_R^2/(\sigma_R^2+1)$, and in this sense balanced the prior evenly over negligible to complete confounding adjustment. The priors for the remaining parameters are the same for the standard model.

Finally, when fitting the semi-parametric CAR model to the lip cancer and covid data in Section \ref{s:lipcancer} and \ref{s:covid} we exploit the default implementation available in {\tt INLA} which runs very fast; this is based on the original parameterization of the semi-parametric, i.e.
$\btheta|\bX\sim\mbox{CAR}
\left(\beta_0{\bf 1}+ \sum_{l=1}^L{\hat \bZ}_lb_l,\sigma_z^2,\lambda_z\right)
$. Priors are as follows: PCP on $\sigma_z^2$, i.e. $\sigma_z\sim\mbox{Exponential}(\xi)$,  with scale parameter $\xi=-\log(0.01)0.31$; PCP on the variance of the adjustment coefficients $\bb$, i.e. $\sigma_b\sim\mbox{Exponential}(\xi)$, with scale parameter  $\xi=-\log(0.01)0.31/0.1$, and $\mbox{logit}(\lambda_z) \sim \mbox{Normal}(0,10)$.

\subsection{Continuous domain}

{\bf Estimation for parsimonious $\Matern$.} The estimation of the conditional model in  (\ref{e:fitted_ygivenx}) simplifies to the estimation of parameters $\{\beta_x, \sigma_z, \sigma, \rho,\nu_z\}$. Priors similar to the common range $\Matern$ model are assigned for $\beta_x$ and $\rho$. We use an exponential prior with rate parameter 1 for the sum ${\sigma_z^2(1-\rho^2) + \sigma^2}$, and a uniform prior for the signal ratio $\frac{\sigma_z^2(1-\rho^2)}{\sigma_z^2(1-\rho^2) + \sigma^2}$. A joint prior is used for $\rho,\nu_z$ to ensure $\rho^2 \le \frac{\nu_{x}\nu_{z}}{(\nu_{x}+\nu_z)/2}$ $\pi(\rho,\nu_z)  = \pi_1(\rho|\nu_z) \pi_2(\nu_z)$ where $\pi_1\sim Unif(-\frac{\sqrt{\nu_{x}\nu_{z}}}{(\nu_{x}+\nu_z)/2},\frac{\sqrt{\nu_{x}\nu_{z}}}{(\nu_{x}+\nu_z)/2})$ and $\pi_2$ is a half Cauchy distribution with mean $\nu_x$ and variance 1000.

\section{Computational details}\label{s:app:comp}

\subsection{Estimation for common range $\Matern$.} To ensure that all parameters are identifiable, we fit the $\Matern$ model of \cite{gneiting2010matern} with common range $\phi_x=\phi_z=\phi_{xz}=\phi$, and additional constraints of $\nu_{xz}>\nu_x$ and $2\nu_{xz}\ge\nu_x+\nu_z$. With these constraints, the maximum for the term $\rho^2\frac{f_{xz}(\bomega)^2}{f_{x}(\bomega)f_{z}(\bomega)}$ in \eqref{e:ygivenx} is obtained at $\bomega = 0$ with value $\rho^2\frac{\nu_{xz}^2}{\nu_x\nu_z}$, hence the $\Matern$ model with the identifiability parameter constraints is valid if $|\rho|<\sqrt{\nu_x\nu_z}/\nu_{xz}$. We use the R package {\tt GpGp} \citep{GpGp} to estimate $\phi_x$, $\nu_x$ and $\sigma_x^2$ by fitting a spatial regression model with $\Matern$ covariance function to the exposure.  The estimated values are plugged into the conditional bivariate $\Matern$ model
\begin{equation}\label{e:fitted_ygivenx}
\bY = \bX\beta_x + \rho\sigma_z R_{zx}{\hat R_x}^{-1}\bX + \bdelta
\end{equation}
where ${\hat R_x}$ is the correlation matrix for $\bX$ computed from the $\Matern$ correlation function with plug-in values of ${\hat\phi_x}$, ${\hat\nu_x}$, and $R_{zx}$ is the cross-correlation matrix between $\bZ$ and $\bX$ determined by parameters $\nu_{xz}$ and the common range $\phi = \hat{\phi_x}$. To reduce computation cost ${\hat\Sigma_x}^{-1}\bX$ can be pre-computed to avoid repetitive evaluation of matrix inversion inside the MCMC algorithm.  
We specify a joint prior for the spatial parameters $\rho$, $\nu_z$ and $\nu_{xz}$, with $\pi(\rho,\nu_z,\nu_{xz}) = \pi_1(\rho | \nu_z,\nu_{xz}) \pi_2(\nu_{z}|\nu_{xz})\pi_3(\nu_{xz})$ where $\pi_1 \sim $ Unif(-$\sqrt{\nu_{x}\nu_{z}}/\nu_{xz}$,$\sqrt{\nu_{x}\nu_{z}}/\nu_{xz}$), $\pi_2 \sim $ Unif(0,$2\nu_{xz}-\nu_x$) and $\pi_3$ is a half Cauchy distribution with mean $\nu_x$ and variance 1000. This specification will ensure that all parameter constrains are met. To complete specification of the hierarchical model, we use an exponential distribution with rate parameter 1 for the sum of variances $\sigma_z^2 + \sigma^2$, a uniform prior for the signal to noise ratio $\frac{\sigma_z^2}{\sigma_z^2 + \sigma^2}$, and a normal prior with mean 0 and variance 100 for the causal effect $\beta_x$. The model is fit using Metropolis-Hastings within Gibbs algorithm with a block update for the spatial parameters. 

\end{document}